\documentclass[%
 reprint,
superscriptaddress,
 amsmath,amssymb,
 aps,
pra,
]{revtex4-2}

\usepackage{setspace}
\usepackage{graphicx}
\usepackage{dcolumn}
\usepackage{bm}
\usepackage[colorlinks=true, allcolors=blue]{hyperref}
\usepackage[export]{adjustbox}
\usepackage{physics}
\usepackage{csquotes}
\usepackage{float}
\usepackage{xcolor}
\usepackage{soul}
\setstcolor{red}
\usepackage{todonotes}
\usepackage{orcidlink}

\begin{document}

\title{Resource-efficient linear-optical generation of GHZ-like states}

\author{Suren A.\,Fldzhyan\,\orcidlink{0000-0002-9174-8019}}
\email[E-mail me at: ]{fldzhyansa@my.msu.ru}
\affiliation{%
 Russian Quantum Center, Bolshoy bul'var 30 building 1, Moscow 121205, Russia
}%
\affiliation{%
 Faculty of Physics, M.\,V. Lomonosov Moscow State University, Leninskie Gory 1, Moscow 119991, Russia
}%

\author{Stanislav S.\,Straupe\,\orcidlink{0000-0001-9810-1958}}
\affiliation{Sber Quantum Technologies Center, Kutuzovski prospect 32, Moscow 121170, Russia}
\affiliation{%
 Russian Quantum Center, Bolshoy bul'var 30 building 1, Moscow 121205, Russia
}%
\affiliation{%
 Faculty of Physics, M.\,V. Lomonosov Moscow State University, Leninskie Gory 1, Moscow 119991, Russia
}%

\author{Mikhail Yu.\,Saygin\,\orcidlink{0000-0001-5494-6801}}
\affiliation{Sber Quantum Technologies Center, Kutuzovski prospect 32, Moscow 121170, Russia}
\affiliation{%
 Faculty of Physics, M.\,V. Lomonosov Moscow State University, Leninskie Gory 1, Moscow 119991, Russia
}%

\begin{abstract}

Heralded multi-photon entanglement generation is a central bottleneck for photonic quantum computing, where resource costs typically skyrocket with target size. We explore efficient methods for generating photon states with tunable entanglement, providing a flexible tool for quantum state engineering. We introduce a theoretical framework that has been numerically validated, demonstrating the capacity to generate GHZ-like states incrementally from non-logical intermediate states. We demonstrate that in certain scenarios -- such as reducing the resource cost for building large maximally entangled GHZ states -- these variable-entanglement states can outperform their fixed-entanglement counterparts. By adjusting intermediate states and optimizing interferometer schemes, we improve photon number cost efficiency of GHZ-like states generation. Our findings indicate that while not a universal solution, non-maximally entangled states offer practical advantages for specific photonic quantum information tasks.

\end{abstract}
\date{\today}

\maketitle

\section{Introduction}\label{sec:Intro}
Photonic quantum information processing is a leading approach to scalable quantum technologies because single photons are naturally suited to low noise transmission, long distance networking, and high fidelity manipulation with mature linear-optical components, detectors, and feedforward electronics. In particular, linear-optical platforms have emerged as a strong candidate for fault-tolerant architectures and for near-term demonstrations of quantum advantage \cite{Aaronson_2013, Zhong_2020, Sund_2024}. Several computational models tailored to photons propagating through interferometers and measured at the output have been developed \cite{Browne_2005, Kieling_2007, Gimeno-Segovia_2015, Bartolucci_2023}, including fusion-based quantum computing (FBQC) \cite{Bartolucci_2023, Bombin_2023, Litinski_2022, Avanesov_2025}.

A central systems level challenge in linear-optical quantum computing is that large entangled resources must be generated probabilistically before or during computation. In FBQC, this challenge is especially acute: the architecture relies on repeatedly preparing smaller entangled photonic states and combining them through fusion operations to build the computational resource. As a result, the practicality of the entire device depends not only on fusion gate design, but also on the efficiency of the resource state factory that supplies entangled states. Improvements at this layer directly affect source requirements, multiplexing overhead, detector load, synchronization complexity, and ultimately the attainable operating regime of a photonic processor.

Entangled photonic states can be generated either through photon-photon interactions mediated by matter systems~\cite{Lim_2009, Gimeno-Segovia_2019, Dhara_2022, Hilaire_2023, weinMinimizingResourceOverhead2025, Petterson_2025, Aqua_2025} or through measurement-induced entanglement using linear optics and postselection \cite{Knill_2001, Knill_2002, Browne_2005}. In this work we focus on the latter route, which remains highly attractive because of its compatibility with integrated interferometric hardware and because it enables heralded state preparation using standard photonic components \cite{Forbes_2025}. Heralding is particularly important in realistic devices, where photon loss and imperfect source synchronization can otherwise propagate errors through subsequent fusion layers.

Among the relevant resource states, GHZ states play a prominent role as a building block for photonic entanglement networks and fusion-based constructions. However, their heralded generation using only linear optics is resource intensive, with overheads that grow rapidly with the target size. Recent progress in numerical optimization of Fock-space transformations has yielded highly efficient few photon protocols \cite{Fldzhyan_2021, Fldzhyan_2023, Gubarev_2020, Gubarev_2021, Chernikov_2023, hartnettAutomatedDiscoveryHeralded2026}, and recent experiments have demonstrated heralded generation of small photonic entangled states \cite{Skryabin_2025, Chen_2024}. At the same time, scalability requires analytical or semi-analytical construction principles that remain tractable as the photon number increases \cite{Bartolucci_2021, Paesani_2021, Pankovich_2024, Bhatti_2025}.

The primate based fusion and bleeding technique introduced by Bartolucci \textit{et al.} \cite{Bartolucci_2021} enables stepwise construction of larger entangled states from smaller resources. These methods expose a structured design space for resource generation, rather than treating each target state as a separate numerical search problem. The bleeding technique is conceptually linked to photon subtraction in quantum optics \cite{Clausen_1999, Bogdanov_2017}, where an input state repeatedly passes through a highly transmissive beamsplitter and a subsequent single photon detection heralds the action of an annihilation operator \cite{Hayrynen_2009, Marek_2018}. However, in their standard form these approaches are tailored to fixed-entanglement intermediates, limiting flexibility when targeting non-maximally entangled states and potentially leaving resource gains unrealized even for maximally entangled targets.

In this paper, we extend the primate fusion framework to the generation of GHZ-like states with tunable entanglement,
\begin{equation}
    |\mathrm{GHZ}_N(s)\rangle=\sqrt{s}\,|10\rangle^N+\sqrt{1-s}\,|01\rangle^N,
\end{equation}
and show that introducing variable entanglement intermediate primates can reduce the average photon resource cost in selected regimes. We quantify efficiency by the average photon cost $\nu$, i.e., the mean number of photons consumed per successful heralded preparation of one target state. This metric is directly relevant for photonic hardware because it captures the burden on single photon sources and multiplexing infrastructure, rather than only the per attempt success probability.

Our main results are twofold. First, for sequential primate fusion, we show that allowing variable entanglement in intermediate primates can improve the photon cost efficiency, including for the preparation of maximally entangled GHZ states in certain ranges of $N$. This identifies a nontrivial design principle: intermediate states that do not themselves exhibit the final target entanglement can be preferable from a resource engineering perspective. Second, we generalize the bleeding procedure by introducing a tunable non-exhaustive bleeding parameter that controls the trade-off between immediate success probability and the quality parameter of the resulting primate. This additional degree of freedom leads to substantial reductions in photon cost, particularly at larger $N$, while remaining compatible with linear-optical implementations.

Beyond the algorithmic optimization itself, the present work provides a fresh physical perspective on measurement-induced entanglement generation: entanglement should be treated as a tunable intermediate resource, not only as a fixed target specification. In our framework, the entanglement parameter of intermediate states and the degree of bleeding exhaustiveness act as coupled control parameters that shape the global resource landscape. This viewpoint may be useful more broadly for designing photonic state factories, where one seeks to optimize device level cost under realistic constraints rather than maximize any single local success probability.

The paper is organized as follows. Section~\ref{sec:Basic descrip} introduces the linear-optical formalism, measurement model, fusion operation, and resource cost metric. Section~\ref{sec:Prim fuse} develops the generalized primate fusion framework and analyzes sequential fusion for generating GHZ-like states. Section~\ref{sec:Bleeding tech} presents the non-exhaustive bleeding protocol and its optimization. Section~\ref{sec:Disc} discusses implications for photonic quantum hardware, FBQC resource generation, and future directions.

\section{Basic Concepts}\label{sec:Basic descrip}

\subsection{Linear-optical evolution}\label{sec:Basic descrip:LO}

The state of photons in the Fock space \cite{Cohen-Tannoudji_2020v3} is described as a superposition of terms involving photon creation operators. For $M$ modes, a pure state takes the form:  
\begin{multline}\label{eq:Fock}
    |\psi\rangle = \sum_{\bm{T}} \alpha_{\bm{T}}|T_1,\cdots, T_M\rangle = \\
    \sum_{\bm{T}} \frac{\alpha_{\bm{T}}}{\sqrt{\prod_{j=1}^{M}T_j!}}\hat{a}^{\dagger T_1}_1\cdots \hat{a}^{\dagger T_M}_M|\text{vac}\rangle,
\end{multline}  
where $|T_1,\cdots, T_M\rangle$ represents the Fock basis with $T_j$ photons in mode $j$.  

Linear optical evolution is governed by a unitary operator $\hat{\mathcal{U}}$, which corresponds to an $M \times M$ interferometer matrix $U$. The evolved state $\hat{\mathcal{U}}|\psi\rangle$ is obtained by appropriately inserting $\hat{\mathcal{U}}\hat{\mathcal{U}}^\dagger$ into each term of the superposition \eqref{eq:Fock}. Under this transformation, the photon creation operators evolve as:  
\begin{equation}
    \hat{\mathcal{U}}\hat{a}^{(\text{in})\dagger}_i\hat{\mathcal{U}}^\dagger = \sum_{j=1}^M \hat{a}^{(\text{out})\dagger}_{j} U_{ji},
\end{equation}  
and the vacuum state remains invariant, $\hat{\mathcal{U}}|\text{vac}\rangle = |\text{vac}\rangle$. These properties simplify state evolution for low photon numbers. For notational simplicity, we will omit explicit (in) and (out) labels where no ambiguity arises.

The interferometer matrix $U$ is constructed by multiplying elementary component matrices. For instance, a beamsplitter with transmittance $t$ is represented by:  
\begin{equation}
    \begin{pmatrix}
        \sqrt{t} & \sqrt{1-t} \\
        \sqrt{1-t} & -\sqrt{t}
    \end{pmatrix},
\end{equation}  
and a phase shifter applying $e^{i\phi}$ to mode $1$ is:  
\begin{equation}
    \begin{pmatrix}
        e^{i\phi} & 0 \\
        0 & 1
    \end{pmatrix}.
\end{equation}  
These components are sufficient to synthesize any unitary interferometer matrix $U$ \cite{Reck_1994}.  

To determine the post-measurement state $|\psi'\rangle$ after photon detection, we apply the measurement projector $\hat{\mathcal{M}}$ to the input state:  
\begin{equation}
    |\psi'\rangle = \hat{\mathcal{M}}|\psi\rangle,
\end{equation}  
where $\||\psi'\rangle\|^2$ gives the probability of the measurement outcome. For example, a measurement of $k$ photons at mode $i$ corresponds to the projector $\hat{\mathcal{M}}^{(k)}_{i} = |k\rangle_i\langle k|_i$. Since photon measurements destroy the detected photons, the projector can be more compactly written as:  
\begin{equation}
    \hat{\mathcal{M}}^{(k)}_{i} = \langle k|_i = \langle\text{vac}|_{i}\frac{\hat{a}^{k}_{i}}{\sqrt{k!}},
\end{equation}  
where $\hat{a}_i$ is the annihilation operator for mode $i$. Here the notation $\langle k|_i$ is understood as the linear functional that selects the $k$-photon component of the state and removes the detected photons --- an operation equivalent to applying the projector $|k\rangle_i\langle k|_i$ and tracing out the measured mode. This form simplifies the notation and is particularly useful when analyzing measurement state transformations.

A practical tool for describing state transformation after measurement is backpropagation. When measuring a state after unitary evolution $\hat{\mathcal{U}}$, the effect of the evolution can be transferred to the measurement operator:
\begin{multline}
    \langle k|_i \hat{\mathcal{U}}|\psi\rangle = \langle\text{vac}|_i\hat{\mathcal{U}}\frac{\left(\hat{\mathcal{U}}^\dagger \hat{a}^{(out)}_i\hat{\mathcal{U}}\right)^k\!\!\!}{\sqrt{k!}}|\psi\rangle=\\\langle\text{vac}|_i\hat{\mathcal{U}}\left[\frac{\left(\sum_{j=1}^{M}U_{ij}\hat{a}^{(in)}_j \right)^k\!\!\!}{\sqrt{k!}}|\psi\rangle\right],
\end{multline}
and then using the algebra of input ladder operators to simplify state for (in) operators before evolving them to (out). This technique simplifies the analysis by partially shifting the complexity of the unitary evolution to the measurement operator, enabling in some cases a more straightforward description of the post-measurement state.

\subsection{Resource cost}
\label{sec:Basic descrip:cost}

We quantify the efficiency of a heralded state generation scheme by the average number of photons consumed per successful production of the target state --- a quantity we call the average photon cost $\nu$.

Consider a linear-optical interferometer where input states $\rho_i$ are combined and a subset of output modes is measured. A specific heralding outcome $\bm{k}$ (e.g., a particular pattern of photon detections) signals successful preparation of the target state $\rho^{(\mathrm{tar})}$ on the remaining modes, and occurs with probability $p^{(\bm{k})}$ \cite{Fldzhyan_2021}.

Each source $i$ has its own average photon cost $\nu^{(i)}$: producing one copy of $\rho_i$ consumes on average $\nu^{(i)}$ photons. By convention, we set $\nu = 1$ for a single photon in a known mode, reflecting that a perfect source consumes exactly one photon per produced state. All other costs are expressed relative to this baseline.

Since each attempt consumes one copy of every input state and attempts are independent, the average number of photons required for one successful generation is

\begin{equation}
\label{eq:cost def}
    \nu^{(\mathrm{tar})} = \frac{\sum_i \nu^{(i)}}{p^{(\bm{k})}}.
\end{equation}

The formalism can also incorporate photon recycling. If a different outcome $\bm{f}$ recovers $r_{\mathrm{ph}}$ recyclable single photons with probability $p^{(\bm{f})}$, the net photon cost reduces to

\begin{equation}\label{eq:cost recycled}
    \nu^{(\mathrm{tar})} = \frac{\sum_i \nu^{(i)} - p^{(\bm{f})} r_{\mathrm{ph}}}{p^{(\bm{k})}}.
\end{equation}

The formalism above assumes ideal conditions: perfect sources, lossless propagation, and ideal detectors. The average photon cost $\nu$ should therefore be interpreted as a fundamental lower bound on resource requirements under these idealizations. In practice, additional overheads from inefficiencies would increase the cost, but the relative comparison between different protocols remains meaningful.

\subsection{Fusion}
\begin{figure}[h]
    \centering
    \includegraphics[width=0.9\linewidth]{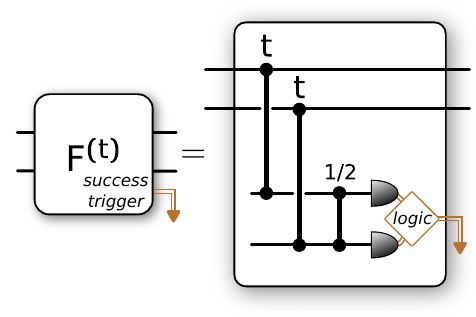}
    \caption{Schematic of the fusion unit. A success trigger can be added to sum detector signals and output a signal when only one photon is detected. Copper wires represent classical signals, with arrows indicating direction. The logical unit discriminates single-photon detection and sends the corresponding signal.}
    \label{fig:0}
\end{figure}

The fusion operation enables the introduction of entanglement between two initially separable photon states. This process involves performing partial measurements on photons from each state to herald the creation of a larger entangled state \cite{Browne_2005}. In our framework, the fusion unit $F^{(t)}$ operates on a pair of modes in the setup depicted in Figure~\ref{fig:0}, where a beamsplitter between two modes is represented by a vertical line with points at the edges. The fusion is considered successful when exactly one photon is measured across the detectors. This process can be fine-tuned by adjusting the beamsplitters' transmittance parameters $t$.

Using the backpropagation outlined in Section~\ref{sec:Basic descrip:LO} we can recover the measurement operators for different cases. The measurement operators $\hat{\mathcal{M}}^{(\text{vac})}_{ij}$ for measuring zero or $\hat{\mathcal{M}}^{(10)}_{ij}$ \& $\hat{\mathcal{M}}^{(01)}_{ij}$ for measuring single photon are:
\begin{equation}\label{eq:meas op 0}
   \hat{\mathcal{M}}^{(\text{vac})}_{ij} = \sqrt{t^{\hat{\mathcal{N}}_{ij}}},
\end{equation}
\begin{equation}\label{eq:meas op 1}
   \hat{\mathcal{M}}^{(10)\text{ or }(01)}_{ij} = \sqrt{\frac{t^{\hat{\mathcal{N}}_{ij}}(1-t)}{2}} \left(\hat{a}_i \pm \hat{a}_j\right),
\end{equation}
where $\hat{\mathcal{N}}_{ij}=\hat{a}^\dagger_i\hat{a}_i + \hat{a}^\dagger_j\hat{a}_j$.

We note that a technique known as boosting can be used to increase the fusion success probability \cite{Gimeno-Segovia_2015, guoBoostedFusionGates2026, Hauser_2025,melkozerovSinglephotonboostedTypeIFusion2026}. While we believe that incorporating boosting could further reduce resource costs, we do not include it in our analysis for the sake of simplicity.

\section{Primate transformation}\label{sec:Prim fuse}

\subsection{Primates}

The work \cite{Bartolucci_2021} introduced the framework of ``primate’’ states --- specialized photonic resource states that enable the stepwise construction of maximally entangled GHZ states via successive fusion operations. Here, we generalize this construction by introducing a broader family of primates. Formally, we define a variable primate state as a pure state of $n+1$ photons distributed across $2n$ modes, denoted $|\pi^{(n)}(\lambda,s)\rangle$, and parameterized by two continuous variables $\lambda$ and $s$. Its explicit form is
\begin{align}\label{eq:primate}
&|\pi^{(n)}(\lambda, s)\rangle = \sqrt{\lambda}|s^{(n)}\rangle + \sqrt{1-\lambda}|0\rangle|\zeta\rangle|0\rangle, \\
&|s^{(n)}\rangle= \sqrt{s}|2\rangle|01\rangle^{n-1}|0\rangle + \sqrt{1-s}|0\rangle|10\rangle^{n-1}|2\rangle,
\end{align}
where $|\zeta\rangle$ is a normalized state of $n+1$ photons in $2n-2$ modes. Informally, $s$ sets the entanglement strength, while $\lambda$ gives the weight of the useful component relative to the junk component. The outermost modes of a primate state are designated for fusion and measurement; the internal modes remain isolated from subsequent interactions.

Our procedure begins with a generation of initial primate states $|\pi^{(1)}(1, s)\rangle = |s^{(1)}\rangle = \sqrt{s}|20\rangle + \sqrt{1-s}|02\rangle$. When $s = 1/2$, the state $\frac{|20\rangle + |02\rangle}{\sqrt{2}}$ can be deterministically created by passing $|11\rangle$ through a balanced beamsplitter. However, generating $|\pi^{(1)}(1, s)\rangle$ from two single photons for arbitrary $s$ is not deterministic \cite{Kieling_2008, De_Gliniasty_2024}.

\begin{figure}[h]
    \centering
    \includegraphics[width=0.5\linewidth]{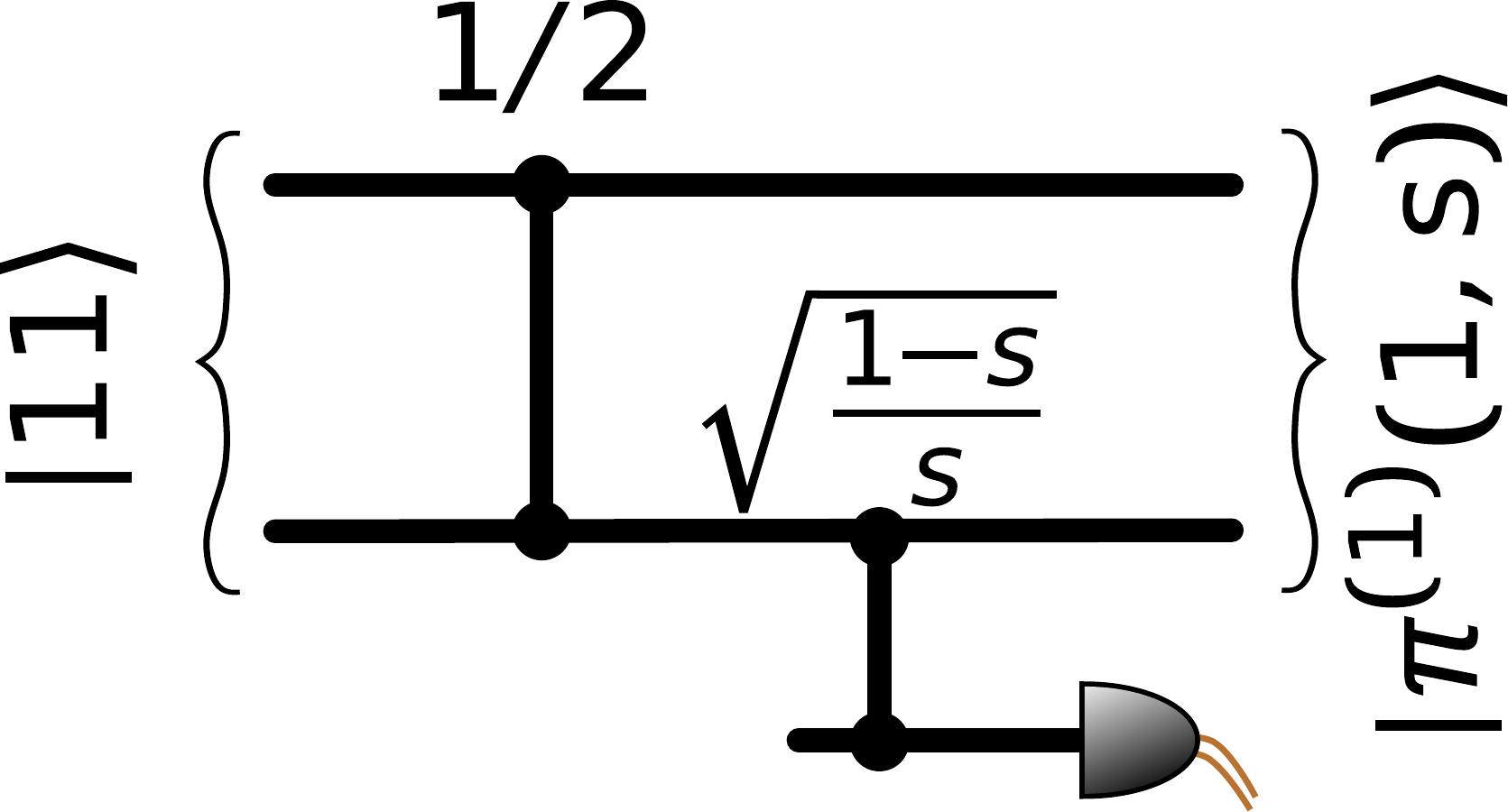}
    \caption{The scheme for generating initial variable primate state $|\pi^{(1)}(1,s)\rangle$.}
    \label{fig:0.4}
\end{figure}

\begin{figure}[h]
    \centering
    \includegraphics[width=0.9\linewidth]{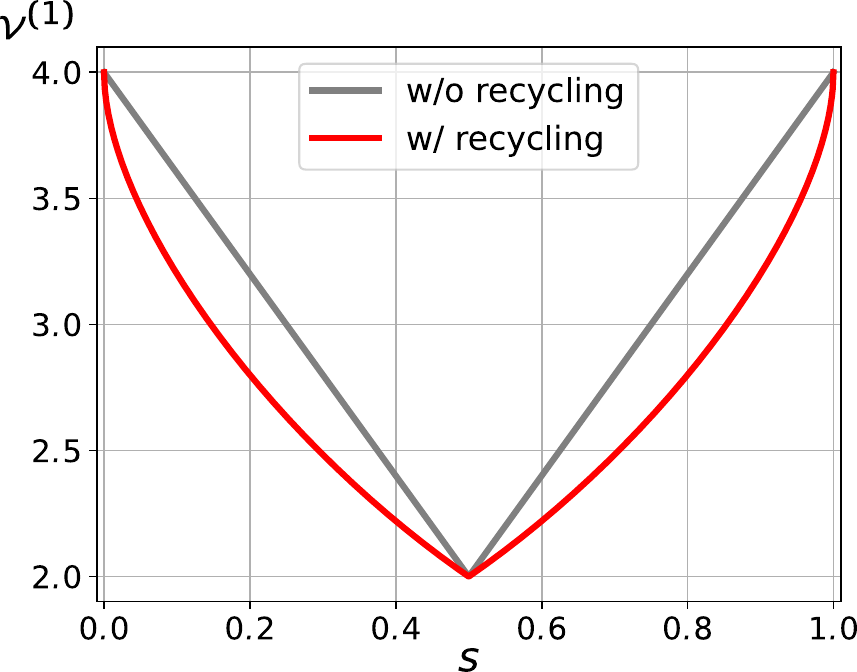}
    \caption{The photon resource cost $\nu^{(1)}$ for generating the initial variable primate state $|\pi^{(1)}(1,s)\rangle$.}
    \label{fig:0.5}
\end{figure}

To generate $|\pi^{(1)}(1, s)\rangle$ we employ the scheme presented in Figure~\ref{fig:0.4}. For $s \geq 1/2$, one can use a beamsplitter between the second mode and an ancilla mode with transmittance $\sqrt{\frac{1-s}{s}}$. After performing a vacuum measurement on the ancilla, the resulting state is:
\begin{equation}
    \frac{|20\rangle + \sqrt{\frac{1-s}{s}}|02\rangle}{\sqrt{2}}=\frac{1}{\sqrt{2s}}|\pi^{(1)}(1,s)\rangle,
\end{equation}
with success probability $\frac{1}{2s}$. Additionally, detecting a single photon in the ancilla mode heralds the presence of a single photon in the second mode, which can be recycled as a resource. The probability of recycling a single photon is given by $\sqrt{\frac{1-s}{s}}\left(1 - \sqrt{\frac{1-s}{s}}\right)$. Using these probabilities and \eqref{eq:cost recycled}, we derive the average photon cost $\nu^{(1)}$ for generating $|\pi^{(1)}(1, s)\rangle$ as:
\begin{equation}
    \nu^{(1)}(s) = 4s - 2\sqrt{1-s}\left(\sqrt{s} - \sqrt{1-s}\right).
\end{equation}
If $s < 1/2$, the process is symmetric under the transformation $s \to 1-s$. This function is illustrated in Figure~\ref{fig:0.5}.

Photon loss can mimic a successful vacuum measurement outcome. For details on loss effects during initial primate generation, see Appendix~\ref{app:sec:init}.

\subsection{Sequential Fusion}

The sequential fusion of primate states results in the generation of a GHZ state. By utilizing the generalized primate state defined in \eqref{eq:primate}, it then becomes possible to generate more general GHZ-like states. We define the target GHZ-like states $|\text{GHZ}_{N}(s)\rangle$ as:  
\begin{equation}
    |\text{GHZ}_{N}(s)\rangle = \sqrt{s}|10\rangle^N + \sqrt{1-s}|01\rangle^N,
\end{equation}  
where qubits are encoded in a dual-rail basis, with single-qubit states represented by the Fock states of a photon distributed across two modes. 

\begin{figure}[h]
    \centering
    \includegraphics[width=0.9\linewidth]{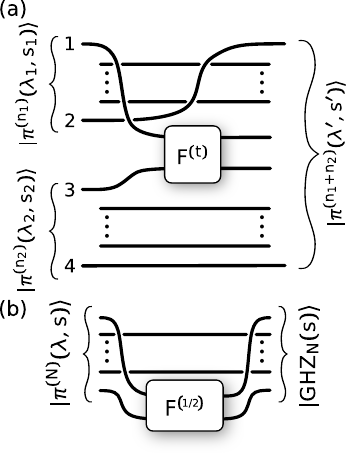}
    \caption{(a) Example of two-primate fusion when the $13$ mode pair is chosen for fusion. (b) The final fusion step that transforms a primate state into a GHZ-like state. Note that the final fusion is always optimal at a beamsplitter transmittance of $1/2$.}
    \label{fig:1}
\end{figure}

Our objective is to generate these GHZ-like states through the fusion of primate states, as illustrated in Figure~\ref{fig:1}. Starting with initial primate states $|\pi^{(1)}(1,s)\rangle$, we sequentially fuse them to construct larger primate states with increasing photon numbers:
\begin{equation}
    |\pi^{(n_1)}(\lambda_1,s_1)\rangle \otimes |\pi^{(n_2)}(\lambda_2,s_2)\rangle \rightarrow |\pi^{(n_1+n_2)}(\lambda',s')\rangle,
\end{equation}
as shown in Figure~\ref{fig:1}a. Subsequently, upon obtaining the primate state with the appropriate photon number $|\pi^{(N)}(\lambda,s)\rangle$, we perform a fusion operation on its outermost modes, as depicted in Figure~\ref{fig:1}b. This final fusion step eliminates the junk component and heralds the successful preparation of the target GHZ-like state.

We now provide explicit formulae for the updated parameters $s'$ and $\lambda'$, as well as the success probability of the fusion process (with detailed derivations in Appendix~\ref{app:Deriv fusion}). The new parameter $s'$ depends on which pair of modes is chosen for fusion. There are two inequivalent cases:
\begin{equation}\label{eq:s pairs}
    s'=
    \begin{cases}
    \frac{s_1s_2}{s_1s_2+(1-s_1)(1-s_2)}, & ij=14\text{ or }23,\\
    \frac{s_1(1-s_2)}{s_1(1-s_2)+(1-s_1)s_2}, & ij=13\text{ or }24,
    \end{cases}
\end{equation}
where the numbering corresponds to the labeling in Figure~\ref{fig:1}a. The success probability $p^{(1)}_{ij}$ (the probability of measuring a single photon) also depends on the pair of modes $ij$ being measured:
\begin{multline}\label{eq:fus succ beg}
    p^{(1)}_{14} = 2t(1-t)\left[s_1\lambda_1 + (1-s_2)\lambda_2 - \right.\\
    \left.2s_1(1-s_2)(1-t^2)\lambda_1\lambda_2 \right],
\end{multline}
\vspace{-3em}
\begin{multline}
    p^{(1)}_{13}=2t(1-t)\left[s_1\lambda_1+s_2\lambda_2-2s_1s_2(1-t^2)\lambda_1\lambda_2\right],\!\!\!\!\!
\end{multline}
\vspace{-3em}
\begin{multline}
    p^{(1)}_{23}=2t(1-t)\left[(1-s_1)\lambda_1+s_2\lambda_2-\right.\\
    \left.2(1-s_1)s_2(1-t^2)\lambda_1\lambda_2\right],
\end{multline}
\vspace{-3em}
\begin{multline}\label{eq:fus succ end}
    p^{(1)}_{24}=2t(1-t)\left[(1-s_1)\lambda_1+(1-s_2)\lambda_2-\right.\\\left.2(1-s_1)(1-s_2)(1-t^2)\lambda_1\lambda_2\right].
\end{multline}
Using these expressions and the fact that the new $\lambda'$ for the state after successful fusion is fraction of the state $|s'^{(n_1+n_2)}\rangle$ probability weight to $p^{(1)}_{ij}$:
\begin{eqnarray}\label{eq:lam prime}
    \lambda'=
    \begin{cases}
    2t(1-t)\lambda_1\lambda_2\frac{s_1s_2+(1-s_1)(1-s_2)}{p^{(1)}_{ij}}, ij=14\text{ or }23\\
    2t(1-t)\lambda_1\lambda_2\frac{s_1(1-s_2)+(1-s_1)s_2}{p^{(1)}_{ij}}, ij=13\text{ or }24
    \end{cases}.
\end{eqnarray}
If the photon cost of generating $|\pi^{(n_k)}(\lambda_k,s_k)\rangle$ is $\nu^{(n_k)}$, then we can derive the average photon cost of generating primate $|\pi^{(n_1+n_2)}(\lambda',s')\rangle$ to be:
\begin{equation}
    \nu^{(n_1+n_2)}=\frac{\nu^{(n_1)}+\nu^{(n_2)}}{p^{(1)}_{ij}},
\end{equation}
which carries on for latter fusions.

The final fusion step, shown in Figure~\ref{fig:1}b, transforms $|\pi^{(N)}(\lambda,s)\rangle$ into the target state $|\text{GHZ}_N(s)\rangle$ with a success probability given by:
\begin{equation}
    p^{(1)} = 2t(1-t)\lambda.
\end{equation}
It is straightforward to show that this fusion achieves its maximum success probability of $\lambda/2$ when the beamsplitter transmittance is set to $t = 1/2$, as explicitly indicated in Figure~\ref{fig:1}b. Consequently, the penultimate fusion, which produces the intermediate state $|\pi^{(N)}(\lambda, s)\rangle$, is most efficient when the product $p^{(1)}_{ij}\lambda/2$ is maximized ($p^{(1)}_{ij}$ being the probability of the penultimate fusion). From \eqref{eq:lam prime} we derive that this condition is also satisfied at $t = 1/2$ (see Appendix~\ref{app:Deriv fusion} for detailed derivation). 

Furthermore, the relationships in \eqref{eq:s pairs}, rewritten in terms of $s'^{-1}-1$, take the form:
\begin{equation}
    (s'^{-1}-1)=
    \begin{cases}
    (s_1^{-1}-1)(s_2^{-1}-1), & ij=14\text{ or }23,\\
    (s_1^{-1}-1)/(s_2^{-1}-1), & ij=13\text{ or }24,
    \end{cases}
\end{equation}
which provides a method for determining one of the very initial states $|\pi^{(1)}(1, s)\rangle$ based on the target state $|\text{GHZ}_N(s)\rangle$.

\begin{figure}[h]
    \centering
    \includegraphics[width=\linewidth]{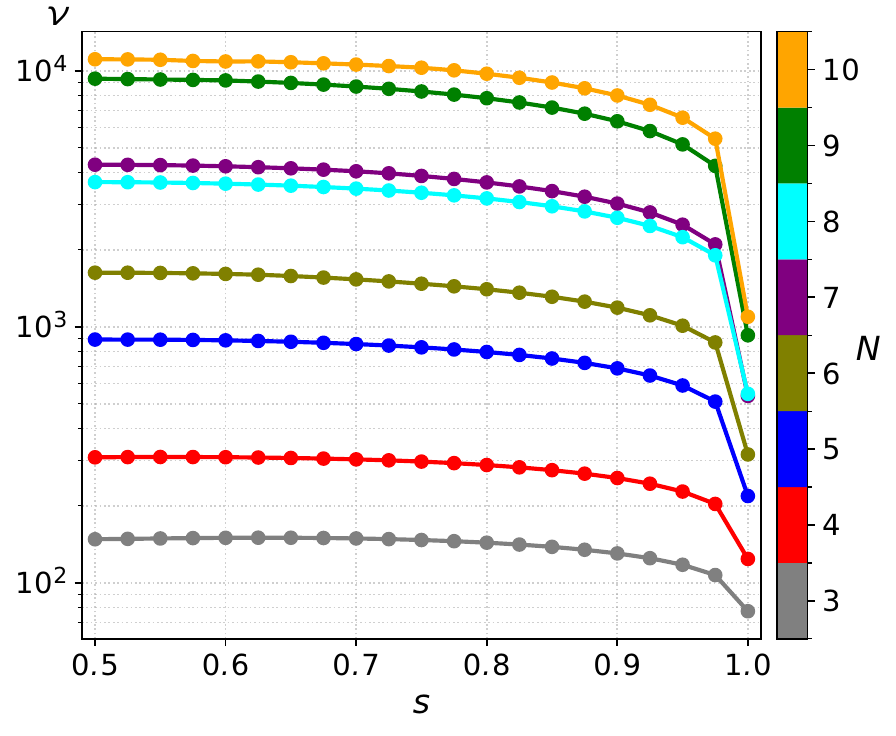}
    \caption{Average photon cost $\nu$ for generating $|\mathrm{GHZ}_N(s)\rangle$ using fusion based primate protocol. Each point represents the minimal cost obtained by optimizing over all addition chains and over the continuous parameters: initial primate entanglement $s$ and beamsplitter transmittances $t$ at fusion units.}
    \label{fig:2}
\end{figure}

When computing single-photon measurement probabilities, we summed the probabilities of the $(01)$ and $(10)$ outcomes for each fusion, even though these outcomes correspond to different resulting states. The rationale for this is that these states can be corrected (via a sign phase shift on a single mode) to yield the same state of the form $|\pi^{(n_1+n_2)}(\lambda', s')\rangle$. However, it is not necessary to actively correct the states during the outlined generation process. If corrections are not applied to the primates after each fusion, the resulting state will still be one that can subsequently be corrected to $|\text{GHZ}_N(s)\rangle$ using phase shift determined by the outcomes of the fusion measurements. The correcting of the state may be avoided altogether by properly adjusting the processing of the fusion network \cite{Bartolucci_2023}.

Now functionally, there are two parameters we can tweak to arrive at the minimal photon resource cost. Those are the parameters $s$ of the initial primates $|\pi^{(1)}(1,s)\rangle$ (which also affect $\nu^{(1)}(s)$) and the transmissions of beamsplitters $t$ at each fusion $F^{(t)}$. But another degree of adaptability we are afforded is the actual path to arrive at the $|\pi^{(N)}(\lambda, s)\rangle$ meaning which primates to fuse and how to fuse them. E.g., the two possible fusion sequences to get $|\text{GHZ}_4\rangle$-like states are: $\{|\pi^{(1)}\rangle\otimes|\pi^{(1)}\rangle\rightarrow|\pi^{(2)}\rangle, |\pi^{(2)}\rangle\otimes|\pi^{(2)}\rangle\rightarrow|\pi^{(4)}\rangle\rightarrow|\text{GHZ}_4\rangle\}$ and $\{|\pi^{(1)}\rangle\otimes|\pi^{(1)}\rangle\rightarrow|\pi^{(2)}\rangle, |\pi^{(2)}\rangle\otimes|\pi^{(1)}\rangle\rightarrow|\pi^{(3)}\rangle, |\pi^{(3)}\rangle\otimes|\pi^{(1)}\rangle\rightarrow|\pi^{(4)}\rangle\rightarrow|\text{GHZ}_4\rangle\}$.

These fusion sequences correspond to addition chains for $N$ \cite{Downey_1981}. An addition chain for $N$ is a sequence of natural numbers where each term is the sum of two previous terms, with the sequence ending in $N$. We focus on a specific type known as star addition chains, in which one of the numbers being summed is always the immediate predecessor \cite{Brauer_1939}. Notably, the minimal addition chain is guaranteed to be found among star chains if $N<12509$ \cite{Knuth_1997}, which comfortably covers the range of values relevant to practical implementations.

Since we lacked a method to determine the provably optimal fusion sequence without exhaustive testing, we considered all possible sequences and optimized the parameters $t$ and $s$ for each sequence individually. For each candidate sequence, we performed a numerical optimization over the continuous parameters: the beamsplitter transmittances $t$ at each fusion step and the entanglement parameters $s$ of the initial primate states. We used a gradient based method (BFGS) to minimize the final average photon cost $\nu$, with multiple random starting points to avoid local minima. Then the best sequence for each target $|\mathrm{GHZ}_{N}(s)\rangle$ was chosen. The results of this procedure are presented in Figure~\ref{fig:2}, where the minimal $\nu$ is plotted against the entanglement parameter $s$ of the target $|\mathrm{GHZ}_{N}(s)\rangle$.

For each $N$, the optimal fusion sequence consistently corresponded to the shortest addition chain (e.g., $\{1,2,4,8\}$ for $N=8$ rather than $\{1,2,3,5,8\}$). This is intuitive: shorter chains involve fewer fusions, reducing the multiplicative build up of photon costs through. In all cases we examined for the fusion‑only protocol, the optimal chain was always among the shortest (i.e., minimal length) addition chains.

From these results, we also observe that using primate states is not the absolute optimal generation strategy for all values of $s$. Naturally, at $s=1$, no primate fusions are necessary because no superposition must be prepared, and the minimal average photon cost reduces to $N$. For other $s$, we cannot guarantee optimality either. The primate construction, while analytically convenient, represents only a restricted subset of heralded schemes achievable with linear optics; more efficient protocols likely exist for any given $s$, though identifying them lies beyond the present scope.

\begin{figure}[h]
    \centering
    \vspace{1em}
    \includegraphics[width=\linewidth]{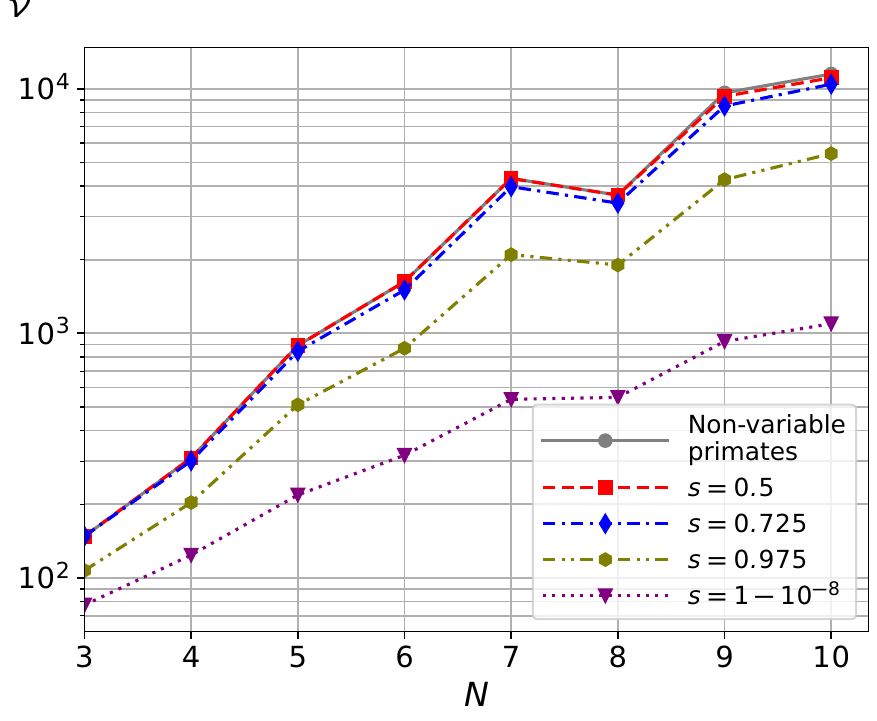}
    \caption{Comparison of resource costs for generating the $|\text{GHZ}_N(s)\rangle$ state. The gray line represents the procedure starting exclusively from $|\pi^{(1)}(1,0.5)\rangle$, while other lines allow for variable initial primates.}
    \label{fig:2.2}
\end{figure}

A particularly noteworthy observation presented in Figure~\ref{fig:2.2} is that even when targeting the maximally entangled state $|\text{GHZ}_N(0.5)\rangle$, employing variable intermediate primate states $|\pi^{(1)}(1,s)\rangle$ yields better performance than using fixed non-variable primates $|\pi^{(1)}(1,0.5)\rangle$ (while still allowing $t$ to vary). This can be explained by the fact that successful fusion probability explicitly depends on the $s_{i}$ (\ref{eq:fus succ beg}-\ref{eq:fus succ end}). The advantage of using variable primates for generating $|\text{GHZ}_N(0.5)\rangle$ starts at $N=7$ and grows with $N$.  For example, at $N=7$ the cost improves by a minuscule amount from $4298.4$ to $4296.0$, and at $N=10$ from $11484.6$ to $11093.5$.

The fact that the resource demand for $N=7$ is greater than for $N=8$ can be partially explained by the fact that the addition chain for $N=7$ is longer: $\{1, 2, 3, 5, 7\}$, compared to $\{1, 2, 4, 8\}$ for $N=8$. Thus more fusions were required, which increased the demand.

\begin{figure}[h]
    \centering
    \includegraphics[width=\linewidth]{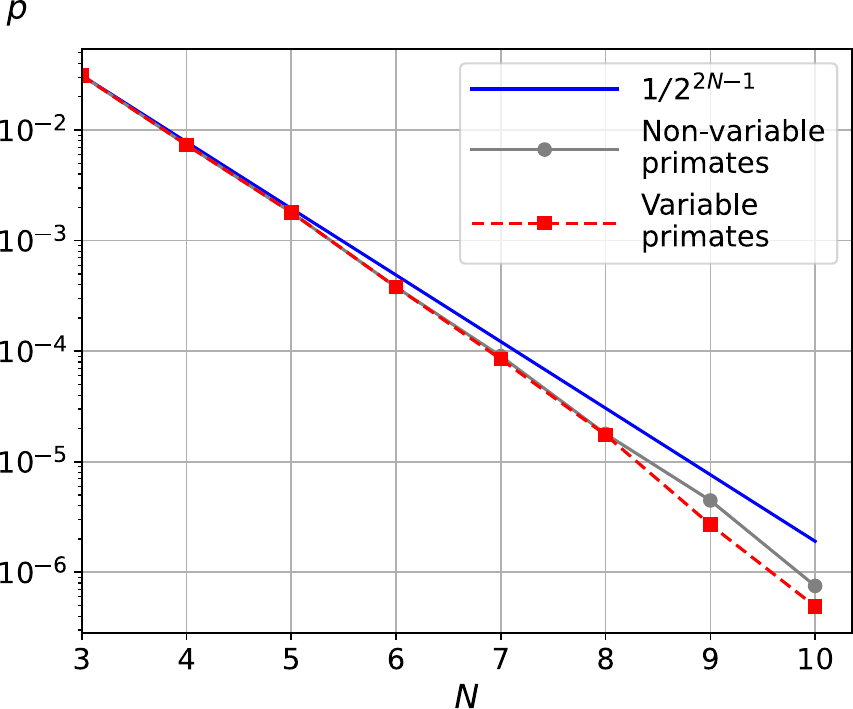}
    \caption{Probability of generating a maximally entangled GHZ state from $2N$ photons in a single pass. Depicted are the maximal achievable probability (which is $1/2^{2N-1}$), along with the probabilities for resource-efficient schemes using both non-variable and variable primates.}
    \label{fig:2.3}
\end{figure}

It is important to distinguish between two key performance metrics. In addition to the average photon number $\nu$ required to generate a single GHZ state, we may also be interested in a probabilistic quantity: namely, the probability of successfully generating a GHZ state from $2N$ photons using the primate-based approach. This corresponds to the single-pass probability that all primate fusions succeed simultaneously. We illustrate this in Figure~\ref{fig:2.3}, alongside the maximal possible single-pass probability for such a scheme, which is $1/2^{2N-1}$ \cite{Gimeno-Segovia_PhD}. This demonstrates that the reduced photon resource cost is achieved at the expense of a lower single-pass success probability.


\section{Bleeding technique}\label{sec:Bleeding tech}
\begin{figure}[h]
    \centering
    \includegraphics[width=\linewidth]{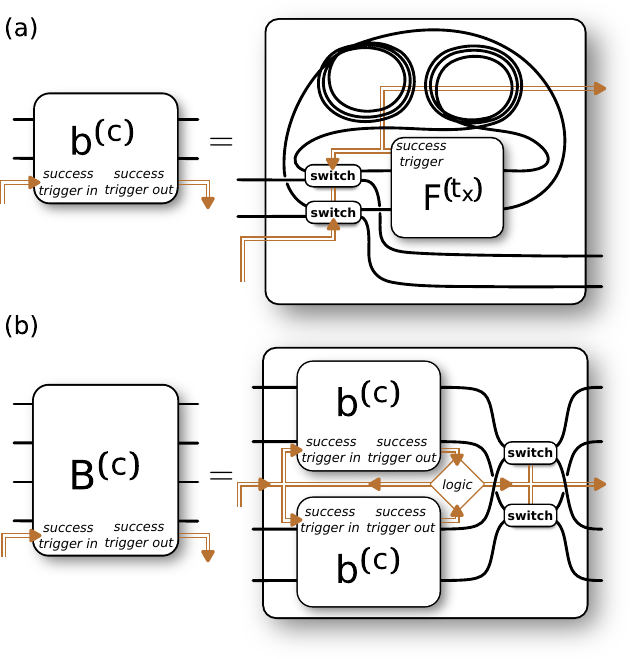}
    \caption{Schematic depiction of the bleeding procedure units $b^{(c)}$ and $B^{(c)}$ is shown in (a) and (b), respectively. The switches in (a) default to passing and switch to crossing upon triggering. The configuration in (b) reroutes channels depending on which subunit $b^{(c)}$ succeeds. The logical unit in (b) must be capable of discriminating between zero, one, or multiple photon measurements and, when necessary, send a triggering signal indicating success or failure terminating the bleeding procedure.}
    \label{fig:2.5}
\end{figure}

Now, there is a more effective way to fuse two primates. Instead of using fusion unit $F^{(t)}$ from Figure~\ref{fig:0} to obtain $|\pi^{(n_1+n_2)}(\lambda',s')\rangle$ from $|\pi^{(n_1)}(\lambda_1,s_1)\rangle$ and $|\pi^{(n_2)}(\lambda_2,s_2)\rangle$ we propose a new unit $B^{(c)}$ based on bleeding \cite{Bartolucci_2021}. A schematic depiction of the bleeding unit $B^{(c)}$ is shown in Figure~\ref{fig:2.5}b. This unit consists of two subunits $b^{(c)}$, depicted in Figure~\ref{fig:2.5}a, which operate simultaneously until one succeeds. Because the success probability per bleeding step is low, processing mode pairs in parallel increases the chance of a successful fusion at the current step. Only two mode pairs can be processed in parallel, since each primate contributes exactly two external modes (superpositions of $|0\rangle$ and $|2\rangle$), giving four modes to pair. These pairs are $14\ \&\ 23$ or $13\ \&\ 24$, with the difference between them affecting the resulting $s'$ (as given in \eqref{eq:s pairs}). 
In principle, the fusion operations $F^{(t_x)}$ should adjust the parameter $t_x$ at each step, which can occur independently of the measurement results. This leads to the characterizing parameter $c$ for $N_b$ step bleeding process, defined as:
\begin{equation}
    0 \leq c = \prod_{x=1}^{N_b} t_x^2 \leq 1.
\end{equation}
Here $c$ quantifies the exhaustiveness of the bleeding process. The limit $c \to 0$ corresponds to exhaustive bleeding, where the protocol continues indefinitely until a photon is detected and the probability for a photon to pass through all beamsplitters vanishes. Conversely, $c > 0$ describes non-exhaustive bleeding, where the process may terminate before photons are measured. For example, given the number of bleeding steps $N_b$ and choosing $t_x = \sqrt{\frac{N_b+1-x}{N_b+2-x}}$, we obtain $c = \frac{1}{N_b+1} \xrightarrow{N_b \to \infty} 0$. In the case of ideal bleeding we assume working in the large number of steps $N_b \gg 1$ regime.

Tunability of $t_x$ is, however, not strictly necessary. If the expected number of bleeding steps $N_b$ is sufficiently large, one can instead choose a constant value, such as $t_x = \sqrt[2N_b]{c}$. In the ideal case of bleeding, the probability of measuring two or more photons becomes vanishingly small due to $t_x \approx 1$ for $c > 0$. For $c = 0$, one can set $t_x = \sqrt[2N_b]{1/N_b}$, which still satisfies $t_x \approx 1$.

Suppose an unsuccessful fusion attempt occurs during the first bleeding step of the pure primate states $|\pi^{(n_1)}(\lambda_1, s_1)\rangle \otimes |\pi^{(n_2)}(\lambda_2, s_2)\rangle$. It is straightforward to show that the state before the second bleeding step becomes $|\pi^{(n_1)}\left(\frac{\lambda_1 t_1^2}{1 - (1-t_1^2)\lambda_1}, s_1\right)\rangle \otimes |\pi^{(n_2)}\left(\frac{\lambda_2 t_1^2}{1 - (1-t_1^2)\lambda_2}, s_2\right)\rangle$, as derived in Appendix~\ref{app:Deriv bleeding}. This indicates that the $\lambda_i$ parameters are updated after the failed attempt. Consequently, the resulting state differs from the original, and we cannot expect the same $\lambda'$ to be achieved in a successful second step as would have been obtained from a successful first step.

To address this issue, we introduce mixed primate states $\pi^{(n)}(\lambda, s)$, defined as:
\begin{equation}\label{eq:pi dens def}
    \pi^{(n)}(\lambda, s) = \lambda |s^{(n)}\rangle\langle s^{(n)}| + (1-\lambda)|0\rangle\langle 0| \otimes \zeta \otimes |0\rangle\langle 0| + \dots.
\end{equation}
Here, $\zeta$ represents the normalized density matrix of an $n+1$ photon state in $2n-2$ modes. The terms denoted by ellipsis correspond to off-diagonal contributions that do not affect probabilities and are eventually ``measured out" during the generation of the target state. This mixed state is not prepared directly, but rather arises from averaging over all possible bleeding histories. The initial primate states themselves remain pure, though they can also be represented in this mixed state form. The updated $\lambda'$ can be calculated as an expectation value over all bleeding steps. This generalization offers a consistent framework for describing the states' evolution under repeated bleeding attempts.

\begin{figure}[h]
    \centering
    \includegraphics[width=0.9\linewidth]{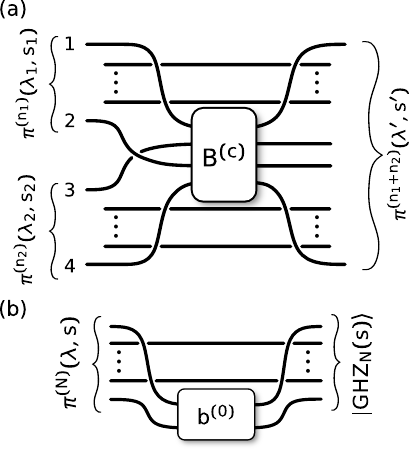}
    \caption{(a) an example of two primate fusion when $13\ \&\ 24$ mode pairs are chosen for interaction on unit $B^{(c)}$. (b) the final unit $b^{(c)}$ that turns a primate into the GHZ-like state. Notice that final $b^{(c)}$ is always optimal at $c=0$.}
    \label{fig:2.8}
\end{figure}

The sequential application of bleeding units is summarized in Figure~\ref{fig:2.8}. The fact that the application of $b^{(0)}$ at $\pi^{(N)}(\lambda,s)$ yields a pure state is obvious from (\ref{eq:meas op 1}, \ref{eq:pi dens def}). In line with the previous section, one chooses the pairs of channels to interact on $B^{(c)}$. The success probability does not depend on $s_i$ and is given by:
\begin{equation}
    p^{(1)}=(1-c)(\lambda_1+\lambda_2-(1-c)\lambda_1\lambda_2).
\end{equation}
If you choose to fuse the mode pairs $14$ and $23$, the resulting transformations are given by:
\begin{eqnarray}
    &&(s'^{-1}-1)=(s_1^{-1}-1)(s_2^{-1}-1),\\
    &&\lambda'=\frac{(s_1s_2+(1-s_1)(1-s_2))\lambda_1\lambda_2(1+c)}{\lambda_1+\lambda_2-(1-c)\lambda_1\lambda_2}.
\end{eqnarray}
Alternatively, if you choose to fuse the mode pairs $13$ and $24$, the corresponding transformations are:
\begin{eqnarray}
    &&(s'^{-1}-1)=(s_1^{-1}-1)/(s_2^{-1}-1),\\
    &&\lambda'=\frac{(s_1(1-s_2)+(1-s_1)s_2)\lambda_1\lambda_2(1+c)}{\lambda_1+\lambda_2-(1-c)\lambda_1\lambda_2}.
\end{eqnarray}
Detailed derivations of these formulae can be found in Appendix~\ref{app:Deriv bleeding}. From these expressions, we can deduce that the success probability of the bleeding process increases as the parameter $c$ decreases. However, a lower value of $c$ leads to a decrease in the parameter $\lambda'$. This trade-off between success probability and the quality of the resulting state (as quantified by $\lambda'$) introduces a non-trivial challenge in the choice of $c$.

Using the derived expressions, we optimize the initial parameters $s$ of $\pi^{(1)}(1, s)$ and the bleeding parameters $c$ to minimize the average photon number cost $\nu$ per $|\text{GHZ}_{N}(s)\rangle$ state generation. Analogous to the process discussed in the previous section, one of the initial primates $\pi^{(1)}(1, s)$ can be determined, and the optimal parameter $c$ is fixed to 0 for specific bleeding units. We fix $c = 0$ for the $B^{(c)}$ unit that fuses two $\pi^{(1)}(1, s)$ states. Similarly, $c = 0$ is fixed for the penultimate $B^{(c)}$ unit that produces the state $\pi^{(N)}(\lambda, s)$, as well as for the last $b^{(c)}$ that generates the desired $|\text{GHZ}_{N}(s)\rangle$ state (see Appendix~\ref{app:Deriv bleeding} for details). These determinations of $c$ simplify the optimization process by reducing the space of variables while maintaining resource efficiency.
\begin{figure}[h]
    \centering
    \includegraphics[width=\linewidth]{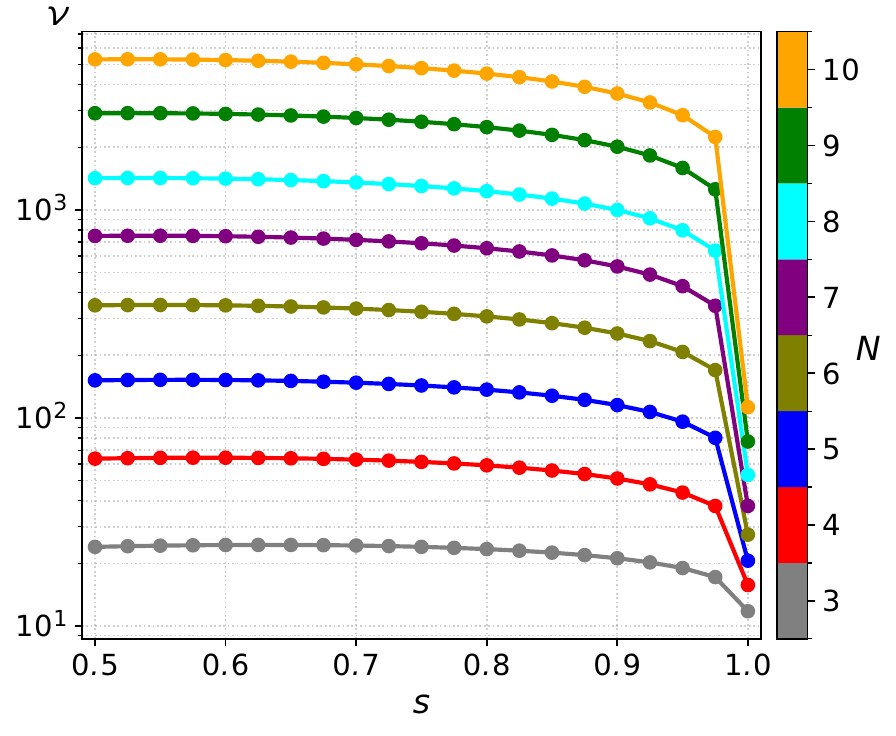}
    \caption{Average photon cost $\nu$ for generating $|\mathrm{GHZ}_N(s)\rangle$ using bleeding based primate protocol. Each point represents the minimal cost obtained by optimizing over all addition chains and over the continuous parameters: initial primate entanglement $s$ and bleeding exhaustiveness $c$ at bleeding units.}
    \label{fig:3}
\end{figure}
\begin{figure}[h]
    \centering
    \vspace{1em}
    \includegraphics[width=\linewidth]{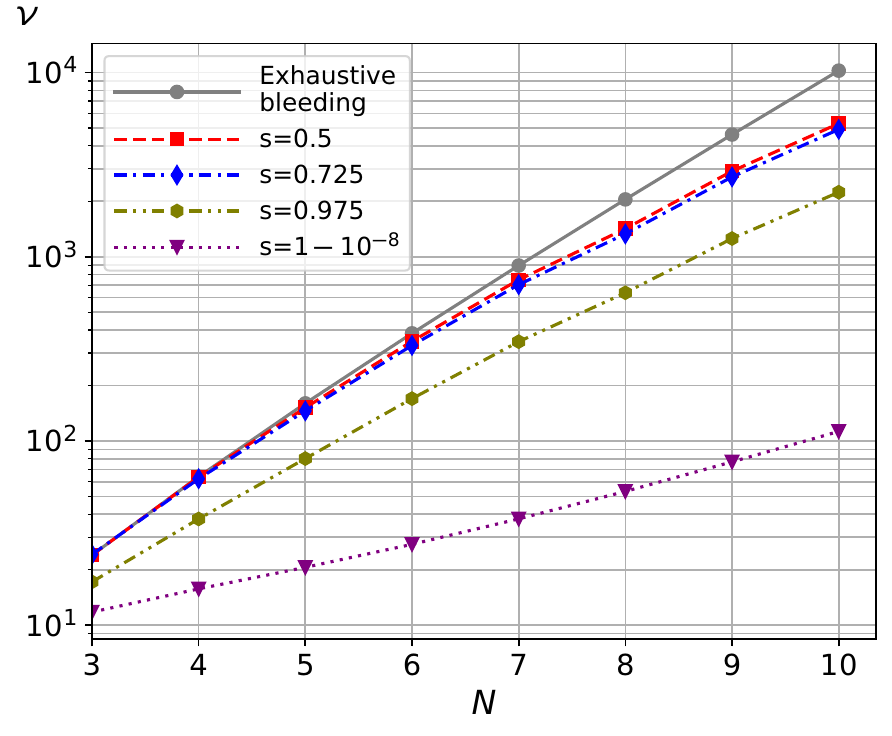}
    \caption{Comparison of resource costs for generating $|\text{GHZ}_N(s)\rangle$ using bleeding. The gray line shows the cost $N2^N$ when initial primates are fixed and only exhaustive bleeding is employed, while other lines allow for variable primates and adjustable $c$ in bleeding units.}
    \label{fig:3.2}
\end{figure}

Figure~\ref{fig:3} shows the average photon cost $\nu$ for generating a single GHZ-like state as a function of the entanglement parameter $s$ and the number of qubits $N$ in the target state. To compare with the value $N2^N$, which represents the average photon cost demonstrated for GHZ state generation using the exhaustive bleeding method in \cite{Bartolucci_2021}, we plot the resource cost dependence on $N$ for different target states in Figure~\ref{fig:3.2}. Our modified bleeding method also shows increasing advantage for generating the maximally entangled state $|\text{GHZ}_N(0.5)\rangle$ as $N$ grows. For example, at $N=4$ the cost improves from $64$ to $63.7$, and at $N=10$ from $10240$ to $5287.6$. In this case, utilizing variable primates does not yield better results when the target state is $|\text{GHZ}_N(0.5)\rangle$, as the optimization converges to non-variable initial primates $|\pi^{(1)}(1,0.5)\rangle$. Nevertheless, significant improvements are still achieved due to the introduced tunability of the bleeding process itself. This means that the introduced non-exhaustive bleeding demonstrates an advantage. 

In contrast to the results in Section~\ref{sec:Prim fuse}, the optimal fusion sequence is not always the shortest. For instance, for $N=4$, the sequence $\{1,2,4\}$ has no variable $c$ and yields a final resource cost of $64$, while the sequence $\{1,2,3,4\}$ has one variable $c$ (with an optimal value of $0.072$) and yields a final resource cost of $63.7$.

\begin{figure}[h]
    \centering
    \includegraphics[width=\linewidth]{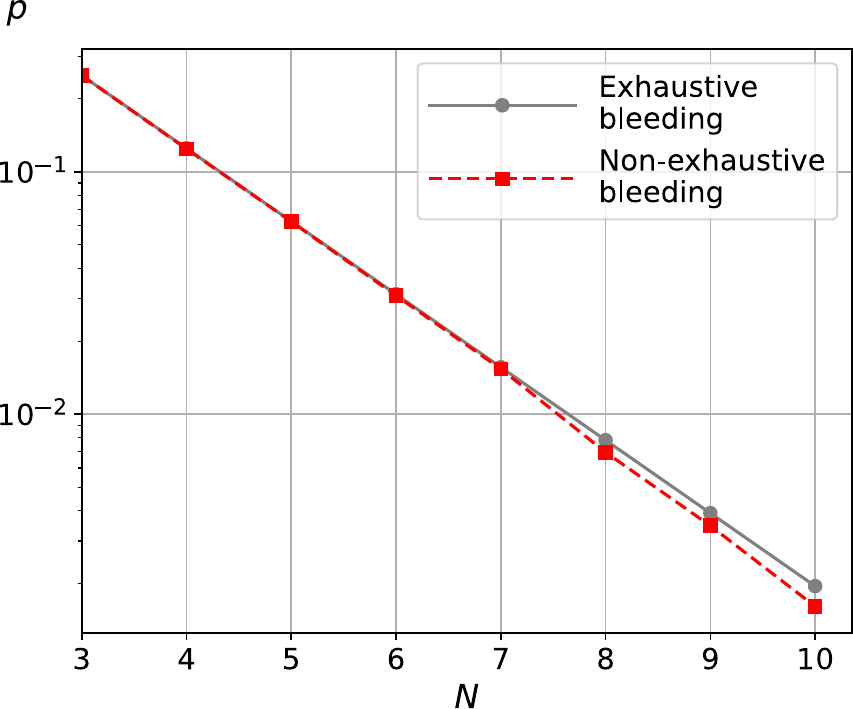}
    \caption{Probability of generating a maximally entangled GHZ state from $2N$ photons in a single pass. Depicted are the probability of exhaustive bleeding (equal to $1/2^{N-1}$), along with the probabilities for the resource-efficient scheme using non-exhaustive bleeding.}
    \label{fig:3.3}
\end{figure}

Again, the comparison of single-pass probability in Figure~\ref{fig:3.3} showcases that the reduced photon resource cost comes at the expense of a lower single-pass success probability. We compare the maximal achievable single-pass probability for an exhaustive bleeding scheme, which is $1/2^{N-1}$ \cite{Bartolucci_2021}, with the single-pass probability of the proposed resource-efficient non-exhaustive bleeding scheme.

\begin{figure}[h]
    \centering
    \includegraphics[width=\linewidth]{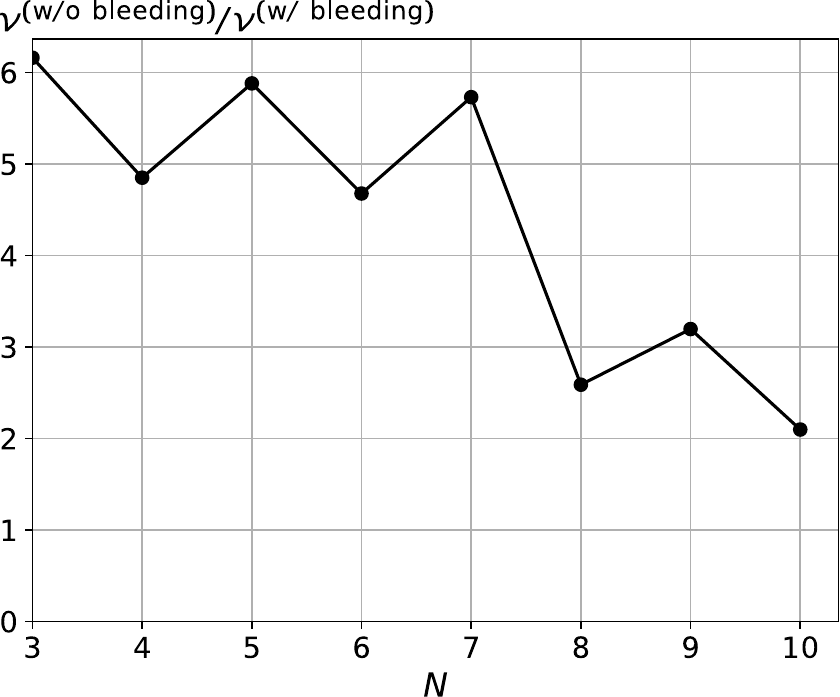}
    \caption{Ratio of the average photon cost using the protocol (Section~\ref{sec:Prim fuse}) to that using the bleeding protocol (Section~\ref{sec:Bleeding tech}) for generating the maximally entangled state $|\mathrm{GHZ}_N(0.5)\rangle$, as a function of $N$. A ratio above 1 indicates an advantage of bleeding.}
    \label{fig:3.4}
\end{figure}

It is noteworthy, when comparing Figure~\ref{fig:2.2} and Figure~\ref{fig:3.2}, that as the number of qubits $N$ increases, the advantage of bleeding for generating $|\text{GHZ}_N(0.5)\rangle$ in comparison to using only fusion diminishes. To better quantify this, Figure~\ref{fig:3.4} showcases the ratio of cost without bleeding to that with bleeding as a function of $N$ for the maximally entangled case $s=0.5$. This confirms that bleeding offers diminishing improvement as $N$ increases. However, even for $N=10$, the bleeding method still provides up to twice the efficiency compared to the approach in Section~\ref{sec:Prim fuse}.

The analysis above assumes an ideal bleeding process with infinitely many steps. In practice, the number of bleeding steps $N_b$ is finite, and optical loss between steps degrades performance. In Appendix~\ref{app:Deriv bleeding}, we extend our formalism to account for both finite $N_b$ and photon loss. Figure~\ref{fig:app:B} shows that for $|\mathrm{GHZ}_5(s)\rangle$, a significant advantage over fusion-only method can be attained even with $N_b=10$ steps even under realistic loss of $0.1,\mathrm{dB}$ per step \cite{liUltracompactUltralowlossBroadband2024,psiquantumteamManufacturablePlatformPhotonic2025}. Thus, while our idealized analysis provides fundamental bounds, the scheme remains practical.


\section{Discussion}\label{sec:Disc}

A central conclusion of this work is that the efficiency of photonic resource state generation can be improved by optimizing the preparation pathway, not only the final target. In our framework, this is done by tuning two internal knobs: the entanglement of intermediate primate states and the exhaustiveness of the bleeding protocol. Optimizing these controls reduces the average photon cost $\nu$ of heralded GHZ-like state generation and therefore directly impacts the resource burden of photonic quantum computing hardware.

This hardware connection is critical for realistic processors, and especially for FBQC-like architectures, where resource states are generated repeatedly and often in parallel. In practice, source brightness and multiplexing efficiency, detector throughput and dead time, feedforward latency, switching network complexity, and loss accumulation across repeated attempts set the dominant constraints. Thus, $\nu$ is not merely a theoretical objective: it serves as a system level proxy for the load imposed on the state factory. Lowering $\nu$ reduces the average number of emitted photons per successful preparation, improving throughput and easing requirements on source arrays and routing infrastructure.

In the sequential fusion setting, we find that variable entanglement intermediate primates can reduce resource cost even when the final target is the maximally entangled state $|\mathrm{GHZ}_N(0.5)\rangle$. This shows that resource-efficient pathways need not preserve maximal entanglement at intermediate stages; instead, the protocol benefits from reshaping entanglement to balance preparation cost against fusion success probabilities. At the same time, the gain in a fusion-only scenario can be modest in some regimes, suggesting a practical design rule: adopting variable entanglement primates should be decided at the platform level, accounting for the added optical control complexity and for imperfections in preparing $|\pi^{(1)}(1,s)\rangle$ at tunable $s$. In integrated implementations, the relevant comparison is a hardware calibrated effective cost that includes propagation loss, mode mismatch, detector inefficiency, and control overhead \cite{Melkozerov_2024, somhorstMitigatingQuantumOperation2026}.

The most immediately actionable experimental outcome is non-exhaustive bleeding. Treating the bleeding exhaustiveness as a tunable parameter $c$ exposes a controllable trade-off between heralding probability and the useful state weight $\lambda$ of the resulting primate. Optimizing this trade-off yields substantial reductions in photon cost relative to exhaustive bleeding. Notably, for maximally entangled GHZ targets the best solutions we find do not require variable entanglement initial primates; the improvement comes from merely tuning the protocol schedule rather than introducing a more demanding initial state primitive, lowering the barrier to implementation. Moreover, although our analysis is formulated through a continuum limit parameter $c$, the same behavior can be approximated with discrete steps, and with nearly constant transmittance per step --- a simplification that can improve calibration robustness and reproducibility in integrated photonics.

Additionally, our results clarify the distinction between maximizing single-pass success probability and minimizing average resource consumption. Resource-efficient strategies may have lower one-shot success probability than protocols optimized for conversion from a fixed photon budget; this reflects an engineering trade-off that depends on whether the primary bottleneck is source rate, detector throughput, switching bandwidth, memory lifetime, or latency. The present framework makes these trade-offs explicit and can be extended naturally to multiobjective optimization and to more general state families and noise models. 

Several directions for future work follow from this study. First, while our focus has been on GHZ-like states, extending this approach to generate states more robust against photon loss is a natural next step. Second, the primate family is unlikely to be globally optimal; broader classes of intermediate states may yield further resource reductions. Third, incorporating practical imperfections --- such as loss, partial distinguishability, detector nonidealities, and feedforward delays --- would transform our idealized cost optimization into a device calibrated design tool. Finally, the tunable state factory approach developed here can be combined with boosted or generalized fusion protocols \cite{Gimeno-Segovia_2015, guoBoostedFusionGates2026, Hauser_2025, melkozerovSinglephotonboostedTypeIFusion2026, Schmidt_2024, rimockGeneralizedTypeII2026, melkozerovEntanglementefficiencyTradeoffsFusionbased2026}, which may reduce the cost further and be especially relevant in fault-tolerant operating regimes. Tunable entanglement GHZ-like states may also find use beyond FBQC; as noted in Ref.~\cite{Zakaryan_2025}, they can be represented as stabilizer states when supplemented with a single ancilla qubit.

In summary, variable entanglement intermediates can reduce photon cost in sequential fusion, and tunable non-exhaustive bleeding can deliver larger, experimentally attractive improvements in heralded GHZ-like state generation. Together, these results strengthen the link between analytical protocol design and the requirements of scalable photonic quantum hardware.

\begin{acknowledgments}
S.A.F. acknowledges the support from the Foundation for the Advancement of Theoretical Physics and Mathematics (BASIS) (Project № 23-2-10-15-1) and the Scholarships of the President of the Russian Federation for postgraduate students.
\end{acknowledgments}

\bibliography{output.bib}

\onecolumngrid
\appendix\label{sec:Appendix}
\section{Fusion}\label{app:Deriv fusion}

In this section we derive formulae from Section~\ref{sec:Prim fuse}. We investigate the fusion of $|\pi^{(n_1)}(\lambda_1,s_1)\rangle\otimes|\pi^{(n_2)}(\lambda_2,s_2)\rangle$. Consider a case when a single photon is measured upon fusion of the $23$ mode pair. Focusing on the $(10)$ measurement and carefully enumerating the terms in the superposition. Starting with the $|s^{(n_1)}\rangle\otimes|s^{(n_2)}\rangle$ term:
\begin{multline}
    \hat{\mathcal{M}}^{(10)}_{23}|s^{(n_1)}\rangle\otimes|s^{(n_2)}\rangle=\\
    \sqrt{\frac{t^{\hat{\mathcal{N}}_{23}}(1-t)}{2}}\left(\hat{a}_2+\hat{a}_3\right)\left(\sqrt{s_1}|2\rangle_1|01\rangle^{n_1-1}|0\rangle_2+\sqrt{1-s_1}|0\rangle_1|10\rangle^{n_1-1}|2\rangle_2\right)\otimes\\
    \otimes\left(\sqrt{s_2}|2\rangle_3|01\rangle^{n_2-1}|0\rangle_4+\sqrt{1-s_2}|0\rangle_3|10\rangle^{n_2-1}|2\rangle_4\right)=\\
    \sqrt{t(1-t)}\left(\left[\sqrt{s_1s_2}|2\rangle_1|01\rangle^{n_1+n_2-1}|0\rangle_4+\sqrt{(1-s_1)(1-s_2)}|0\rangle_1|10\rangle^{n_1+n_2-1}|2\rangle_4\right]+\right.\\
    \left.\sqrt{t^2(1-s_1)s_2}|0\rangle_1|10\rangle^{n_1-1}\left(|12\rangle_{23}+|21\rangle_{23}\right)|01\rangle^{n_2-1}|0\rangle_4\right).
\end{multline}
The term in square brackets is proportional to the state $|s'^{(n_1+n_2)}\rangle$, with the updated parameter:
\begin{equation}
    s'=\frac{s_1s_2}{s_1s_2+(1-s_1)(1-s_2)}.
\end{equation}
The other types of terms are:
\begin{multline}
    \hat{\mathcal{M}}^{(10)}_{23}|s^{(n_1)}\rangle\otimes|0\rangle_3|\zeta_2\rangle|0\rangle_4=\\
    \sqrt{\frac{t^{\hat{\mathcal{N}}_{23}}(1-t)}{2}}(\hat{a}_2+\hat{a}_3)\left(\sqrt{s_1}|2\rangle_1|01\rangle^{n_1-1}|0\rangle_2+\sqrt{1-s_1}|0\rangle_1|10\rangle^{n_1-1}|2\rangle_2\right)|0\rangle_3|\zeta_2\rangle|0\rangle_4=\\
    \sqrt{t(1-t)(1-s_1)}|0\rangle_1|10\rangle^{n_1}|\zeta_2\rangle|0\rangle_4,
\end{multline}
\vspace{-3em}
\begin{multline}
    \hat{\mathcal{M}}^{(10)}_{23}|0\rangle_1|\zeta_1\rangle|0\rangle_2\otimes|s^{(n_2)}\rangle=\\
    \sqrt{\frac{t^{\hat{\mathcal{N}}_{23}}(1-t)}{2}}(\hat{a}_2+\hat{a}_3)|0\rangle_1|\zeta_1\rangle|0\rangle_2\left(\sqrt{s_2}|2\rangle_3|01\rangle^{n_2-1}|0\rangle_4+\sqrt{1-s_2}|0\rangle_3|10\rangle^{n_2-1}|2\rangle_4\right)=\\
    \sqrt{t(1-t)s_2}|0\rangle_1|\zeta_1\rangle|01\rangle^{n_2}|0\rangle_4.
\end{multline}
To compute the success probability $p^{(1)}$, we sum the probabilities of all terms, appropriately weighted by $\lambda_1$ and $\lambda_2$. Accounting for both the $(10)$ and $(01)$ cases (which only differ by the sign in the superposition) with a factor of 2, we find:
\begin{multline}
    p^{(1)}=2t(1-t)\left(\lambda_1\lambda_2\left(s_1s_2+(1-s_1)(1-s_2)+2t^2(1-s_1)s_2\right)+\lambda_1(1-\lambda_2)(1-s_1)+(1-\lambda_1)\lambda_2s_2\right)=\\
    2t(1-t)\left(\lambda_1(1-s_1)+\lambda_2s_2-2(1-t^2)(1-s_1)s_2\lambda_1\lambda_2\right),
\end{multline}
and the $\lambda'$ is determined by the ratio of the probability weight of the $|s^{(n_1+n_2)}\rangle$ state to the total success probability:
\begin{equation}\label{eq:app:lam fus}
\lambda'=\frac{t(1-t)\lambda_1\lambda_2\left(s_1s_2+(1-s_1)(1-s_2)\right)}{p^{(1)}/2}.
\end{equation}
By following a similar procedure, we can derive analogous expressions for any other fused mode pairs.

\subsection{Optimizing the final fusions}
Regarding the final fusion step, we derive

\begin{multline}\label{eq:app final fus}
\hat{\mathcal{M}}^{(10)}_{12}|s^{(N)}\rangle = \sqrt{\frac{t^{(\mathrm{last})\hat{\mathcal{N}}_{12}}(1-t^{(\mathrm{last})})}{2}}(\hat a_1+\hat a_2)\left(\sqrt{s}|2\rangle_1|01\rangle^{N-1}|0\rangle_2+\sqrt{1-s}|0\rangle_1|10\rangle^{N-1}|2\rangle_2\right)\\
= \sqrt{t^{(\mathrm{last})}(1-t^{(\mathrm{last})})}\left(\sqrt{s}|10\rangle^N + \sqrt{1-s}|01\rangle^N\right)=\sqrt{t^{(\mathrm{last})}(1-t^{(\mathrm{last})})}|\mathrm{GHZ}_N(s)\rangle.
\end{multline}

Taking into account both the measurement operators $\hat{\mathcal{M}}^{(10)}_{12}$ and $\hat{\mathcal{M}}^{(01)}_{12}$ and the probability $\lambda$ that the state contains the $|s^{(N)}\rangle$ component, the total success probability for the final fusion is

\begin{equation}
p^{(\mathrm{last})} = 2t^{(\mathrm{last})}(1-t^{(\mathrm{last})})\lambda,
\end{equation}
which is maximal at $t^{(\mathrm{last})}=1/2$. Consequently, the average photon cost for generating the GHZ state is

\begin{equation}
\nu^{(\mathrm{GHZ}_N)} = \frac{\nu^{(N)}}{2t^{(\mathrm{last})}(1-t^{(\mathrm{last})})\lambda} = \frac{\nu^{(n_1)}+\nu^{(n_2)}}{2t^{(\mathrm{last})}(1-t^{(\mathrm{last})})\lambda p^{(\mathrm{penult})}},
\end{equation}

where the second equality assumes the penultimate fusion combined states $|\pi^{(n_1)}\rangle$ and $|\pi^{(n_2)}\rangle$ with success probability $p^{(\mathrm{penult})}$. To minimize the resource cost, one must therefore maximize the product $\lambda p^{(\mathrm{penult})}$ by appropriately tuning the penultimate fusion. Examining Eq.~\eqref{eq:app:lam fus}, we see that

\begin{equation}
    \nu^{(\mathrm{GHZ}_N)} \propto \frac{1}{t^{(\mathrm{last})}(1-t^{(\mathrm{last})})\,t^{(\mathrm{penult})}(1-t^{(\mathrm{penult})})},
\end{equation}

where $t^{(\mathrm{penult})}$ denotes the beamsplitter transmittance used in the penultimate fusion. This product is maximized when both transmittances are set to $t = 1/2$, thereby minimizing the resource cost.

\section{Bleeding}\label{app:Deriv bleeding}

In this section we derive formulae in Section~\ref{sec:Bleeding tech}. We begin by considering a product state of two primates, denoted as $\pi^{(n_1)}(\lambda_1, s_1) \otimes \pi^{(n_2)}(\lambda_2, s_2)$, where each primate is defined as:

\begin{equation}
    \pi^{(n_i)}(\lambda_i, s_i) = \lambda_i |s^{(n_i)}_i\rangle\langle s^{(n_i)}_i| + (1-\lambda_i)|0\rangle\langle 0| \otimes \zeta_i \otimes |0\rangle\langle 0| + \dots.
\end{equation}

To reiterate: $\zeta_i$ represents the normalized density matrix of an $n_i+1$ photon state in $2n_i$ modes, while the terms denoted by ellipsis correspond to off-diagonal contributions that do not influence probabilities. For simplicity, we omit ellipsis in subsequent derivations.

Let $t_x$ denote the fusion units' parameters at the $x$-th bleeding steps. As discussed in the main text, we propose performing the bleeding operation simultaneously on two pairs of modes --- for instance, the innermost and outermost pairs $23$ and $14$. After a single bleeding step, there are two key measurement outcomes of interest: the projection operator for the case where zero photons are measured (given by \eqref{eq:meas op 0}) and the case where one photon is measured (given by \eqref{eq:meas op 1}). Since a single bleeding unit $B^{(c)}$ consists of two subunits $b^{(c)}$, which operate simultaneously, we must apply an additional zero-photon measurement operator for each case to account for the evolution of the state.

\subsubsection{Case 1: No Photon Detected}

If no photons are detected during the $(x+1)$-th bleeding step, the resulting state is:
\begin{multline}\label{eq:app:bleed_vac}
    \hat{\mathcal{M}}^{(\text{vac})}_{14} \hat{\mathcal{M}}^{(\text{vac})}_{23} \pi^{(n_1)}\left(\lambda_1(x), s_1\right) \otimes \pi^{(n_2)}\left(\lambda_2(x), s_2\right) \hat{\mathcal{M}}^{\dagger(\text{vac})}_{14} \hat{\mathcal{M}}^{\dagger(\text{vac})}_{23} = \\
    \left(\lambda_1(x) t_{x+1}^2 |s^{(n_1)}_1\rangle\langle s^{(n_1)}_1| + (1-\lambda_1(x))|0\rangle\langle 0|_1 \otimes \zeta_1 \otimes |0\rangle\langle 0|_2\right)\\ \otimes \left(\lambda_2(x) t_{x+1}^2 |s^{(n_2)}_2\rangle\langle s^{(n_2)}_2| + (1-\lambda_2(x))|0\rangle\langle 0|_3 \otimes \zeta_2 \otimes |0\rangle\langle 0|_4\right),
\end{multline}
where $\lambda_i(x)$ denotes the parameter values after the $x$-th unsuccessful step, with $\lambda_i(0) = \lambda_i$.
The probability of this outcome is:
\begin{equation}\label{app:p zero}
p^{(0)}_{x+1} = \left(1 - (1-t_{x+1}^2)\lambda_1(x)\right)\left(1 - (1-t_{x+1}^2)\lambda_2(x)\right),
\end{equation}
where $p^{(0)}_x$ denotes the probability that no photons have been measured in the first $x$ steps. The updated parameters $\lambda_1(x+1)$ and $\lambda_2(x+1)$ are then:
\begin{equation}\label{app:lam prog}
    \lambda_i(x+1) = \frac{\lambda_i(x) t_{x+1}^2}{1 - (1-t_{x+1}^2)\lambda_i(x)},
\end{equation}
These expressions can be simplified. The probability of no detection up to and including the $x$-th step is:
\begin{equation}
p^{(0)}(\leq x) = \left(1 - \left(1-\prod_{x'=1}^{x}t_{x'}^2\right)\lambda_1\right)\left(1 - \left(1-\prod_{x'=1}^{x}t_{x'}^2\right)\lambda_2\right),
\end{equation}
and the updated parameters after the $x$-th step are:
\begin{equation}\label{app:lam prog to 0}
    \lambda_i(x) = \frac{\lambda_i \prod\limits_{x'=1}^{x}t_{x'}^2}{1 - \left(1-\prod\limits_{x'=1}^{x}t_{x'}^2\right)\lambda_i}.
\end{equation}

\subsubsection{Case 2: Single Photon Detected}

We now consider the case where a single photon is detected in one of the four modes after propagation through the fusion units. We fix the bleeding step $x$ at which this measurement occurs. Specifically, suppose a $(10)$ outcome is observed for the $23$ mode pair, while vacuum is detected on $14$. The effect of this measurement on the pure state $|s^{(n_1)}_1\rangle |s^{(n_2)}_2\rangle$ is given by:

\begin{multline}\label{app:meas}
    \hat{\mathcal{M}}^{(\text{vac})}_{14}\hat{\mathcal{M}}^{(10)}_{23}|s^{(n_1)}_1\rangle |s^{(n_2)}_2\rangle = \sqrt{\frac{t^{\hat{\mathcal{N}}_{14}+\hat{\mathcal{N}}_{23}}_x(1-t_x)}{2}}\left(\hat{a}_2+\hat{a}_3\right)\cdot\\
    \cdot\left(\sqrt{s_1}|2\rangle_1|01\rangle^{n_1-1}|0\rangle_2 + \sqrt{1-s_1}|0\rangle_1|10\rangle^{n_1-1}|2\rangle_2\right)\left(\sqrt{s_2}|2\rangle_3|01\rangle^{n_2-1}|0\rangle_4 + \sqrt{1-s_2}|0\rangle_3|10\rangle^{n_2-1}|2\rangle_4\right)=\\
    \sqrt{t^3_x(1-t_x)}\left(\left[\sqrt{s_1s_2}|2\rangle_1|01\rangle^{n_1+n_2-1}|0\rangle_4+\sqrt{(1-s_1)(1-s_2)}|0\rangle_1|10\rangle^{n_1+n_2-1}|2\rangle_4\right]+\right.
    \\\left.+\sqrt{(1-s_1)s_2}|0\rangle_1|10\rangle^{n_1-1}\left(|21\rangle_{23}+|12\rangle_{23}\right)|01\rangle^{n_2-1}|0\rangle_4\right).
\end{multline}

The state in square brackets is proportional to the new state $|s'^{(n_1+n_2)}\rangle$, where
\begin{equation}
    s'=\frac{s_1s_2}{s_1s_2+(1-s_1)(1-s_2)} \quad \Rightarrow \quad \left(s'^{-1}-1\right)=\left(s_1^{-1}-1\right)\left(s_2^{-1}-1\right).
\end{equation}

There are four possible single-photon measurement outcomes from fusing the $14$ and $23$ pairs, all of which can be corrected via simple phase shifts or mode permutations to yield the same final state structure. From \eqref{app:meas}, the weight $\lambda'_x$ of the resulting $|s'^{(n_1+n_2)}\rangle$ state after step $x$ is:
\begin{equation}\label{app:lam prime}
    \lambda'_x = 4\lambda_1(x-1)\lambda_2(x-1) t_x^3(1-t_x)\frac{s_1s_2+(1-s_1)(1-s_2)}{p^{(1)}_x},
\end{equation}
where the factor $\lambda_1(x-1)\lambda_2(x-1)$ arises from the probability of the component $|s^{(n_1)}_1\rangle\langle s^{(n_1)}_1|\otimes|s^{(n_2)}_2\rangle\langle s^{(n_2)}_2|$ in the input $\pi_1\otimes\pi_2$, and $p^{(1)}_x$ is the total probability of detecting a single photon at step $x$, including contributions from terms involving $\zeta_1$ or $\zeta_2$. For completeness, we provide the expression for $p^{(1)}_x$:
\begin{multline}
    p^{(1)}_{x}=2t_x(1-t_x)\left(\lambda_1(x-1)+\lambda_2(x-1)-2(1-t^2_x)\lambda_1(x-1)\lambda_2(x-1)\right)=\\
    =\frac{2t_x(1-t_x)\prod\limits_{x'=1}^{x-1}t_{x'}^2\left(\lambda_1+\lambda_2-2\left(1-\prod\limits_{x'=1}^{x}t_{x'}^2\right)\lambda_1\lambda_2\right)}{p^{(0)}(\leq x-1)}.
\end{multline}

\subsection{Continuum Approximation}

At its core, the bleeding process assumes that $t_x \approx 1$ for most steps. To simplify the analysis, we approximate the discrete transmission steps with a continuous representation, letting $t_x \approx 1 - \Delta(x)dx\approx e^{-\Delta(x)dx}$. Using this approximation, we focus only on terms up to the first power of $dx$. This allows us to derive from \eqref{app:lam prog} differential equations describing the evolution of $\lambda_1(x)$ and $\lambda_2(x)$:
\begin{equation}
    \lambda_i(x+dx) = \lambda_i(x)(1-2\Delta(x)dx)(1+2\Delta(x)dx\lambda_i(x)) \quad \Rightarrow \quad \dot{\lambda}_i(x) = -2\Delta(x)\lambda_i(x)(1-\lambda_i(x)),
\end{equation}
with the solution:
\begin{equation}
    \lambda_i(x) = \frac{1}{1+\frac{1-\lambda_i(0)}{\lambda_i(0)}e^{2\int\limits^x_0 \Delta(x')dx'}}=\frac{\lambda_i e^{-2\int\limits_0^{x}\Delta(x')dx'}}{1-\left(1-e^{-2\int\limits_0^{x}\Delta(x')dx'}\right)\lambda_i},
\end{equation}
which equals \eqref{app:lam prog to 0} if $e^{-2\int_0^{x}\Delta(x')dx'}=\prod_{x'=1}^{x}t_{x'}^2$.
The probability of not measuring any photons at step $x$ from \eqref{app:p zero} can be approximated to be:
\begin{equation}
    p^{(0)}(x) = e^{-2\Delta(x)\left(\lambda_1(x)+\lambda_2(x)\right)dx},
\end{equation}
and the cumulative probability of not measuring any photons up to step $x$ is:
\begin{equation}
    p^{(0)}(< x) = e^{-2\int\limits^{x}_0\Delta(x')\left(\lambda_1(x')+\lambda_2(x')\right)dx'}=\left(1-\left(1-e^{-2\int\limits_0^{x}\Delta(x')dx'}\right)\lambda_1\right)\left(1-\left(1-e^{-2\int\limits_0^{x}\Delta(x')dx'}\right)\lambda_2\right).
\end{equation}
The value of $\lambda'(x)$ at step $x$ is approximated from \eqref{app:lam prime} to be:
\begin{equation}
    \lambda'(x) = 4\lambda_1(x)\lambda_2(x)\Delta(x)dx\frac{s_1s_2+(1-s_1)(1-s_2)}{dp^{(1)}(x)},
\end{equation}
where it is emphasized that probability of measuring single photon $dp^{(1)}(x)$ in the vicinity of $x$ is proportional to $dx$.

Our main interest lies in determining $\lambda'$ for the final state after the entire bleeding process. Due to the nature of the density matrix formalism, this reduces to an expectation value:
\begin{equation}
    \lambda' = \frac{1}{p^{(1)}}\int\limits^{\infty}_0 \lambda'(x)p^{(0)}(< x)dp^{(1)}(x) = \frac{4\left(s_1s_2+(1-s_1)(1-s_2)\right)}{p^{(1)}}\int\limits^{\infty}_0 \lambda_1(x)\lambda_2(x)\Delta(x)p^{(0)}(< x)dx.
\end{equation}
With bleeding success probability $p^{(1)}$ given by:
\begin{equation}
    p^{(1)}=1-p^{(0)}(<\infty),
\end{equation}
because probability of measuring $2$ or more photons is of higher order on $dx$. Remarkably, these integrals can be evaluated analytically, yielding the key results:
\begin{align}
    & (s'^{-1}-1) = (s_1^{-1}-1)(s_2^{-1}-1), \\
    & p^{(1)}=(1-c)\left(\lambda_1+\lambda_2-(1-c)\lambda_1\lambda_2\right),\label{eq:app p bleed}\\
    & \lambda' = \left(s_1s_2+(1-s_1)(1-s_2)\right)\frac{\lambda_1\lambda_2(1+c)}{\lambda_1+\lambda_2-(1-c)\lambda_1\lambda_2},\label{eq:app lam bleed} \\
    & c = e^{-2\int\limits^{\infty}_0 \Delta(x)dx} \approx \prod_{x=1}^{N_b}t^2_x,
\end{align}
where we have used $\lambda_i=\lambda_i(0)$. The parameter $c$ represents the strength of the bleeding procedure: the closer $c$ is to $0$, the more exhaustive the bleeding is. Alternatively, if we fuse pairs $13$ and $24$ instead of $14$ and $23$, the expressions change:
\begin{align}
    & (s'^{-1}-1) = (s_1^{-1}-1)/(s_2^{-1}-1), \\
    & \lambda' = \left(s_1(1-s_2)+(1-s_1)s_2\right)\frac{\lambda_1\lambda_2(1+c)}{\lambda_1+\lambda_2-(1-c)\lambda_1\lambda_2}.
\end{align}
With the help of the fact that $\frac{\lambda_1+\lambda_2}{2\lambda_1\lambda_2} \geq 1$, it can be easily proven that $\frac{dp^{(1)}}{dc} \leq 0$ and $\frac{d\lambda'}{dc} \geq 0$.  Consequently, optimizing $c$ requires balancing these competing effects to minimize the overall resource cost.

We have used a continuum approximation because it yielded a neat single parameter $c$ that governs a perfect bleeding behavior with infinite number of steps and transmissions arbitrarily close to $1$. For reference here are the exact expressions when no such assumptions are made:
\begin{align}
    &p^{(1)}=\sum_{x=1}^{N_b}2t_x(1-t_x)\left(\lambda_1+\lambda_2-2\left(1-\prod_{x'=1}^{x}t_{x'}^2\right)\lambda_1\lambda_2\right)\prod_{x'=1}^{x-1}t_{x'}^2,\label{eq:app:finite bleed prob}\\
    &\lambda'=4\lambda_1\lambda_2\frac{s_1s_2+(1-s_1)(1-s_2)}{p^{(1)}}\sum_{x=1}^{N_b}t_x^3(1-t_x)\prod_{x'=1}^{x-1}t_{x'}^4,\label{eq:app:finite bleed lam}
\end{align}
where summation goes up to the number of bleeding steps $N_b$. If other mode pairs are fused then $s_1(1-s_2)+(1-s_1)s_2$ is substituted instead of $s_1s_2+(1-s_1)(1-s_2)$.

\subsection{Optimizing initial and final bleeding stages}

We first consider fusing two elementary primates $|\pi^{(1)}(1,s_1)\rangle$ and $|\pi^{(1)}(1,s_2)\rangle$. Setting $\lambda_1=\lambda_2=1$ in~(\ref{eq:app p bleed}) and~(\ref{eq:app lam bleed}) gives
\begin{equation}
    p^{(1)}=(1-c^2),\qquad \lambda' = s_1s_2 + (1-s_1)(1-s_2),
\end{equation}
so $c=0$ is always optimal when fusing initial primates.

For the final fusion, a $b^{(c)}$ unit acts on $\pi^{(N)}(\lambda, s)$. When no photon is detected at step $x+1$, we have
\begin{equation}
    \hat{\mathcal{M}}^{(\mathrm{vac})}_{12}\pi^{(N)}(\lambda(x), s)\hat{\mathcal{M}}^{\dagger(\mathrm{vac})}_{12}= \lambda(x)t^{2}_{x+1}|s^{(N)}\rangle\langle s^{(N)}| + (1-\lambda(x))|0\rangle\langle 0|_1\otimes \zeta\otimes|0\rangle\langle 0|_2,
\end{equation}
where $\lambda(x)=\frac{\lambda(0)\prod\limits_{x'=1}^{x}t^2_{x'}}{p^{(0)}(\leq x)}$ and $p^{(0)}(\leq x)=1-\left(1-\prod\limits_{x'=1}^{x}t^2_{x'}\right)\lambda(0)$ is the probability of measuring zero photons up to step $x$. When a single photon is detected,~(\ref{eq:app final fus}) yields
\begin{align}
    \hat{\mathcal{M}}^{(10)}_{12}\pi^{(N)}(\lambda(x), s)\hat{\mathcal{M}}^{\dagger(10)}_{12}= \lambda(x)t_{x+1}(1-t_{x+1})|\mathrm{GHZ}_N(s)\rangle\langle\mathrm{GHZ}_N(s)|,
\end{align}
with total probability $p^{(1)}(x+1)=2\lambda(x)t_{x+1}(1-t_{x+1})$ (accounting for both $\hat{\mathcal{M}}^{(10)}_{12}$ and $\hat{\mathcal{M}}^{(01)}_{12}$). Summing contributions from all steps,
\begin{multline}\label{eq:app:p last bleed}
    p^{(1)}=\sum_{x=0}^{\infty}p^{(1)}(x+1)p^{(0)}(\leq x)=2\sum_{x=0}^{\infty}\lambda(x)t_{x+1}(1-t_{x+1})p^{(0)}(\leq x)
    =2\lambda(0)\sum_{x=0}^{\infty}t_{x+1}(1-t_{x+1})\prod^{x}_{x'=1}t^2_{x'}\\
    \approx 2\lambda(0)\int^{\infty}_{0} \Delta(x)\,e^{-2\int^{x}_{0}\Delta(x')dx'}dx
    = \lambda(0)\left(1-e^{-2\int^{\infty}_{0}\Delta(x)dx}\right)=\lambda(0)(1-c^{(\mathrm{last})}).
\end{multline}

As in Appendix~\ref{app:Deriv fusion}, we can similarly analyze the penultimate bleeding unit $B^{(c^{(\mathrm{penult})})}$. If it fuses states $\pi^{(n_1)}$ and $\pi^{(n_2)}$, the resource cost becomes
\begin{equation}
    \nu^{(\mathrm{GHZ}_N)}=\frac{\nu^{(N)}}{\lambda(1-c^{(\mathrm{last})})}=\frac{\nu^{(n_1)}+\nu^{(n_2)}}{\lambda p^{(\mathrm{penult})} (1-c^{(\mathrm{last})})}.
\end{equation}
Examining ~(\ref{eq:app p bleed}) and~(\ref{eq:app lam bleed}), we find
\begin{equation}
    \nu^{(\mathrm{GHZ}_N)}\propto\frac{1}{(1-c^{2(\mathrm{penult})})\,(1-c^{(\mathrm{last})})},
\end{equation}
so the resource cost is minimized when $c^{(\mathrm{penult})}=c^{(\mathrm{last})}=0$.

\subsection{Bleeding imperfection}
\begin{figure}
\centering
\includegraphics[width=0.5\linewidth]{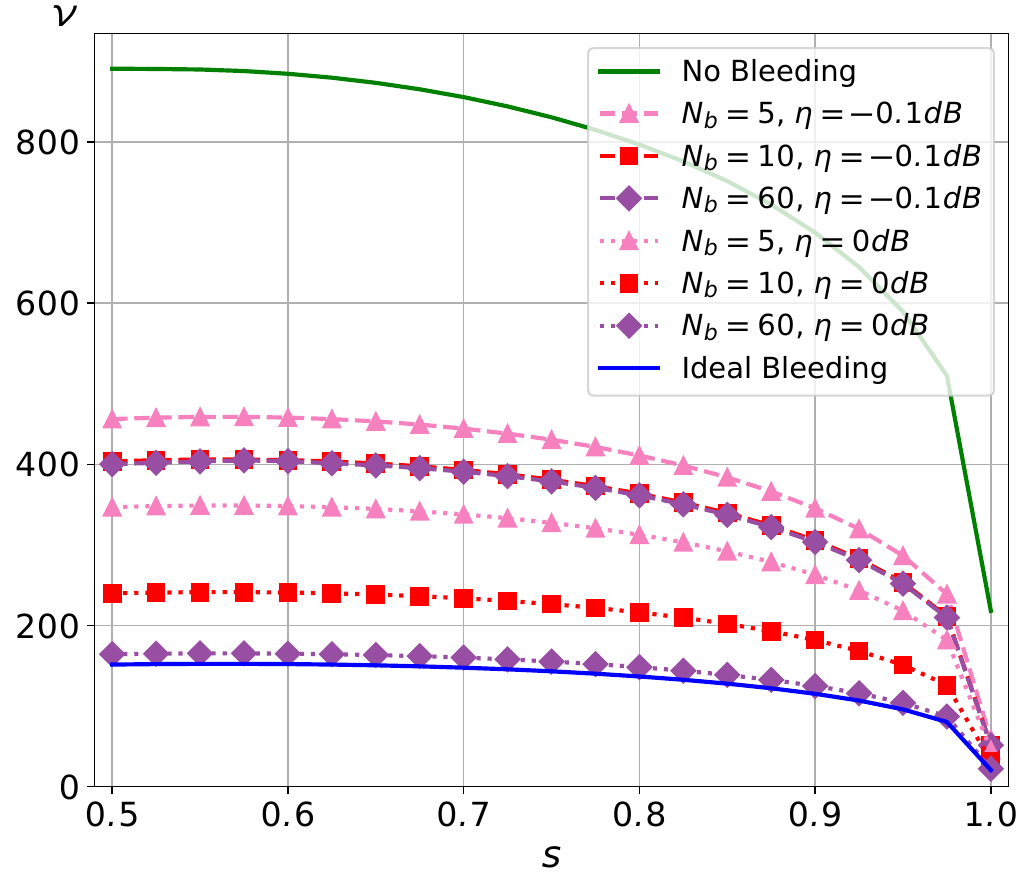}
\caption{ Photon cost $\nu$ for generating $|\mathrm{GHZ}_5(s)\rangle$ under nonideal bleeding conditions. The green curve shows the cost without bleeding (fusion-only). The blue curve corresponds to ideal bleeding with infinitely many steps and no loss. Other curves show realistic scenarios with a finite number of steps $N_b$ and loss $\eta$. Even with finite steps and realistic loss, the cost remains significantly below the no-bleeding case, demonstrating the practicality of the scheme.
}
\label{fig:app:B}
\end{figure}

Using \eqref{eq:app:finite bleed prob} and \eqref{eq:app:finite bleed lam}, we can account for a finite number of bleeding steps by optimizing over each $t_x$ individually rather than using a single aggregated parameter $c$. Incorporating optical loss between bleeding steps is also straightforward, though it introduces imperfect heralding. That being said, the lossy component falls outside the dual-rail encoding and can in principle be accounted for downstream. To model loss, we modify Eq.~\eqref{eq:app:bleed_vac} by including a loss channel for the fused mode pairs and postselecting on the outcome where no photon is lost:
\begin{multline}
    \Phi_{\mathrm{no~photon~lost}}\left[\hat{\mathcal{M}}^{(\text{vac})}_{14} \hat{\mathcal{M}}^{(\text{vac})}_{23} \pi^{(n_1)}\left(\lambda_1(x), s_1\right) \otimes \pi^{(n_2)}\left(\lambda_2(x), s_2\right) \hat{\mathcal{M}}^{\dagger(\text{vac})}_{14} \hat{\mathcal{M}}^{\dagger(\text{vac})}_{23}\right] = \\
    \left(\lambda_1(x) t_{x+1}^2 \eta^2 |s^{(n_1)}_1\rangle\langle s^{(n_1)}_1| + (1-\lambda_1(x))|0\rangle\langle 0|_1 \otimes \zeta_1 \otimes |0\rangle\langle 0|_2\right) \\
    \otimes \left(\lambda_2(x) t_{x+1}^2\eta^2 |s^{(n_2)}_2\rangle\langle s^{(n_2)}_2| + (1-\lambda_2(x))|0\rangle\langle 0|_3 \otimes \zeta_2 \otimes |0\rangle\langle 0|_4\right),
\end{multline}
where $\eta$ is the probability that a photon survives without loss. With this modification, \eqref{eq:app:finite bleed prob} and \eqref{eq:app:finite bleed lam} become:
\begin{align}
    &p^{(1)}=\sum_{x=1}^{N_b}2t_x(1-t_x)\left(\lambda_1+\lambda_2-2\left(1-\prod_{x'=1}^{x}t_{x'}^2\eta^2\right)\lambda_1\lambda_2\right)\prod_{x'=1}^{x-1}t_{x'}^2\eta^2,\\
    &\lambda'=4\lambda_1\lambda_2\frac{s_1s_2+(1-s_1)(1-s_2)}{p^{(1)}}\sum_{x=1}^{N_b}t_x^3(1-t_x)\prod_{x'=1}^{x-1}t_{x'}^4\eta^4,
\end{align}
with analogous expressions for the other choice of mode pairs. And for the probability of last fusion \eqref{eq:app:p last bleed}
\begin{equation}
    p^{(1)}=2\lambda\sum_{x=1}^{N_b}t_{x}(1-t_{x})\prod^{x-1}_{x'=1}t^2_{x'}\eta^2.
\end{equation}
Figure~\ref{fig:app:B} illustrates the impact of finite bleeding steps and loss on the generation of $|\mathrm{GHZ}_5(s)\rangle$, demonstrating that near optimal performance can be achieved with a modest number of steps even in the presence of realistic loss. The figure also reveals that loss imposes a fundamental limitation on bleeding: beyond a certain point, increasing the number of steps $N_b$ yields diminishing returns and does not significantly reduce the resource cost.

\section{Imperfect distillation}\label{app:sec:init}
If a photon is lost in the ancilla mode before measurement, the resulting vacuum detection can mimic a successful heralding event. The loss of $k$ photons in channel $i$ is modeled by the Kraus operator

\begin{equation}
    \hat{\mathcal{K}}^{(k)}_{i}=\sum_{n=0}^{\infty}\sqrt{\binom{n}{k}\eta^{n-k}(1-\eta)^k}|n-k\rangle\langle n |_i=\sqrt{\eta^{\hat{\mathcal{N}}_i}(1-\eta)^k}\frac{\hat{a}^k_i}{\sqrt{k!}},
\end{equation}
where $\eta$ is the probability that a photon survives. Denoting the beamsplitter transmittance by $t=\sqrt{\frac{1-s}{s}}$, the state before the detectors in Fig.~\ref{fig:0.4} is
\begin{equation}
    |\phi\rangle=\frac{|200\rangle+t|020\rangle+\sqrt{2t(1-t)}|011\rangle+(1-t)|002\rangle}{\sqrt{2}}.
\end{equation}
Applying the loss channel followed by vacuum measurement on the third mode yields

\begin{equation}
    \langle \mathrm{vac}|_3\left(\sum_{k=0}^{2}\hat{\mathcal{K}}^{(k)}_3|\phi\rangle\langle\phi|\hat{\mathcal{K}}^{\dagger(k)}_3\right)|\mathrm{vac}\rangle_3=\frac{1}{2s}|s^{(1)}\rangle\langle s^{(1)}|+\alpha |01\rangle\langle 01|+\beta|00\rangle\langle 00|,
\end{equation}
where $\alpha$ and $\beta$ depend on $s$ and the loss magnitude. The $\beta$ component can be absorbed into our existing formalism as part of the junk component. The $\alpha$ component, however, may propagate through the scheme undetected, ultimately yielding $|00\rangle$ in one of the $|\mathrm{GHZ}_N(s)\rangle$ qubits rather than the intended dual-rail $|01\rangle$ or $|10\rangle$ effectively manifesting as logical level photon loss after state generation. Protocols for managing such loss in downstream applications have been developed, for instance, in the context of fusion-based quantum computing \cite{Melkozerov_2024, bartolucci2025comparisonschemeshighlyloss}. For simplicity we assume that no such loss is present.


\end{document}